\documentclass[useAMS,usenatbib,babel,superscriptaddress]{mnras}

\usepackage[english]{babel}
\usepackage[tbtags]{amsmath}
\usepackage{amssymb,amsfonts,textcomp}
\usepackage{array}
\usepackage{supertabular}
\usepackage{hhline}
\usepackage{hyperref}
\usepackage{graphicx}
\usepackage{color}
\definecolor{Red}{rgb}{0.65,0.08,0.05}

\providecommand{\sorthelp}[1]{}

\newcommand{\ve}[1]{\mathbf{#1}}

\newcommand{\Epsilon}{\mathcal{E}}
\DeclareMathOperator{\Tr}{Tr}

\title[Clustering and alignments of peak patch  haloes]{Statistical exploration of halo anisotropic clustering and intrinsic alignments 
with the mass-Peak Patch algorithm}

\begin{document}


\author[
B. Regaldo-Saint Blancard et al.]{
\parbox[t]{\textwidth}{Bruno Regaldo-Saint Blancard$^{1}$, Sandrine Codis$^{2,3,4}$\thanks{E-mail: codis@iap.fr}, J. Richard Bond$^{5}$, George Stein$^{6}$
}
\vspace*{6pt}\\
\noindent$^{1}$ Laboratoire de Physique de 
l'ENS, ENS,
Universit\'e PSL, CNRS, Sorbonne Universit\'e, Universit\'e de Paris,
75005 Paris, France
\\
\noindent$^{2}$ Institut d'Astrophysique de Paris, CNRS and Sorbonne Universit\'e, UMR 7095, 98 bis Boulevard Arago, 75014 Paris, France\\
\noindent$^{3}$ Institut de Physique Th\'eorique, Universit\'e Paris-Saclay, CNRS, CEA, 91191 Gif-sur-Yvette, France\\
\noindent$^{4}$ AIM, CEA, CNRS, Universit\'e Paris-Saclay, Universit\'e Paris Diderot, Sorbonne Paris Cit\'e, 91191 Gif-sur-Yvette, France\\
\noindent$^{5}$ Canadian Institute for Theoretical Astrophysics, 60 St. George St, Toronto, Canada, ON M5S 3H8\\
\noindent$^{6}$ Berkeley Center for Cosmological Physics,
341 Campbell Hall, University of California, Berkeley, CA 94720, USA
}

\maketitle


\begin{abstract}
The anisotropy or triaxiality of massive dark matter haloes largely defines the structure of the cosmic web, in particular the  filaments that join the haloes together. Here we investigate such oriented correlations in mass-Peak Patch halo catalogues by using the initial strain tensor of spherical proto-halo regions to orient the haloes. To go beyond the spherically averaged two-point correlation function of haloes we use oriented stacks to compute oriented two-point correlations: we explicitly break isotropy by imposing a local frame set by the strain tensor of the reference halo before stacking neighbouring haloes. Beyond the exclusion zone of the reference halo, clustering is found to be strongly enhanced along the major direction of the strain tensor as expected. This anisotropic clustering of haloes along filaments is further quantified by using a spherical harmonics decomposition. Furthermore,
we compute the evolution of cluster-scale halo principal directions relative to those of their neighbours and show that there are strong correlations extending up to very large scales. In order to provide calculations more suitable to observational confrontations, we also utilize 2D projected versions of some equivalent correlation functions. Finally, we show that the multipole structure of the mass-peak patch halo's anisotropic clustering can be qualitatively captured in an analytic treatment based on peak theory.
Though highly informative, giving the same qualitative features as the oriented correlations found from the simulation catalogue, analytic evaluation involves extensive use of Monte Carlo methods, which is also what the simulated catalogue uses, taking into account as they do the adaptive nature of the mass-peak patch mass hierarchy and all non-local complexities associated with the exclusion of smaller haloes overlapping with larger ones: there is no substitute for the mass-Peak Patch simulation-based determination of oriented and anisotropic correlations. 
\end{abstract}

\section{Introduction}

Investigating the distribution of matter at scales ranging from a few megaparsecs to hundreds of megaparsecs has been a very prolific field of Cosmology for the last fifty years. From the first theoretical works on superclustering and observations in the seventies through current large galaxy surveys, our knowledge of large scale structures has grown enormously, with much more to come with the upcoming massive galaxy surveys. The interconnectivity in the distribution of matter on the largest scales is  referred to as the \textit{Cosmic Web} \citep{klypin&shandarin83,bkp96,pogosyanetal1998,2008LNP...740..335V}, the natural nonlinear evolution of primordial density fluctuations through gravitational instability in an expanding universe, giving rise to the galaxy clusters, filaments, sheets and bubble-like cosmic voids that we see today. The present day large-scale structure of the Universe was already imprinted in the anisotropies of the initial conditions: initial peaks \citep[hereafter BBKS]{bardeen/bond/etal:1986} later form massive clusters at the nodes of the cosmic web, correlation bridges in between collapse to form filaments \citep{bkp96} surrounded by walls and voids in this fully connected cosmic network.

Observations and numerical simulations complement one another to get a deeper understanding of our Universe. But still technical issues coming from their manipulation are rather different. Information from observational data is subject to experimental measurement error and is restricted to the past light cone constraint. Simulations, though limited by the theoretical approximations used and computer performance, can be used to derive mock catalogues of all aspects of the matter distribution, including precise positions and velocities and field configuration properties for identified objects, over all space and time, not just on the past light cone. The mass-Peak Patch computational halo-finding methodology described in \cite{ref1} (hereafter BM1) and other works of that era are accurate on the past light cone as well as over all space and time, and allow rapid calculations of ensembles of such realizations, as parameters defining the cosmology vary. Though traditional N-body simulations are more accurate, to this day N-body ensembles suitable for mocking all manner of massive observational campaigns remain limited. The mass-Peak Patch algorithm was further improved in \cite{2019MNRAS.483.2236S} for massively parallel computation. 
It allows  identification from an initial (linear) state distribution of matter of regions (contiguous mass-patches) that will collapse to form virialized objects in the final state space, haloes. Other approximate methods of varying speeds have  been put forward over the past decades. e.g.,  Pinocchio \citep{2002MNRAS.331..587M} is one such example. The attention to fast simulation techniques has been driven by a dire need to address very costly statistical problems such as likelihood surface construction for large galaxy surveys, for which covariance matrices encode the simplest characterization, e.g.,  \citep[see e.g][and references therein]{2016Galax...4...53M,2019MNRAS.482.1786L}. 

Here, we do not focus on these observationally-related problems but rather undertake a theoretical exploration, using such approximate methods to determine the clustering of haloes in the web, in particular focussing on the question of halo alignments. 
Peak patch simulations can address this since they use coarse-grained ellipsoidal dynamics of the halo patches, following the nonlinear development of initial anisotropic tidal forces which correlate from halo to halo. Triaxial collapse along the last axis to equilibrate is  slower than in isotropic spherical collapse for highly initially-sheared patches. 

These tidal forces   {which dictate the formation sites of the filamentary cosmic web \citep{1993ApJ...418..544V,bkp96,1996MNRAS.281...84V}}, also shape the morphology of haloes notably through tidal stretching \citep{Catelan01}  and spin them up through tidal torques \citep[see e.g][for a review]{lee&pen00,Cri++01,schaefer09}. Because tides are set on large scales by the cosmic web, a large-scale coherence of the shapes and spins of galaxies and haloes is naturally expected \citep[see for instance][for a modelling of the spin alignments of galaxies induced by the cosmic web]{ATTT}.
Various numerical works \footnote{  {We refer to \cite{2018MNRAS.473.1195L} for a review of algorithms designed to extract the cosmic web from simulations.}} have indeed demonstrated that the shape of haloes tend to align with filaments and walls in dark matter simulations \citep{2006ApJ...652L..75P,2007MNRAS.375..184B,hahnetal07,calvoetal07}, a result also found more recently for galaxies in cosmological hydrodynamical simulations \citep{2018MNRAS.481.4753C}. This is in agreement with observations which have evidenced a clear radial alignment of massive clusters \citep{2012MNRAS.423..856S,2017MNRAS.468.4502V}, observations that go way back to the eighties.

These halo and galaxy alignments are key to understand galaxy formation and evolution notably through the role of the environment versus internal processes. They also represent a worrisome contamination for current and, even more so, for future large photometric surveys which are designed to map the distribution of matter through measurement of  cosmic shear, e.g., KiDS, DES, HSC, CFIS, Euclid, LSST \citep[see e.g][for recent reviews]{Joachimi15,2015SSRv..193..139K,Kiessling15,troxel15}. Indeed, intrinsic alignments of galaxies induce a spurious signal in the lensing-induced shear-shear correlation function which biases significantly the cosmological analysis if not accounted for properly \citep{Kirk10,Krause15}. Different strategies have been developed over the course of the last few years, from perturbative approaches \citep{2019PhRvD.100j3506B,2020JCAP...01..025V} to semi-analytical models \citep{Schneider10,joachimi13a,Joachimi13b,2020arXiv200302700F}, and now  state-of-the-art hydrodynamical computations can address this, by simulating large cosmological volumes while keeping enough resolution for galaxies \citep{codis14,Tenneti15a,Chisari16,Velliscig15b,Hilbert16}. One of the major difficulties in modelling such an effect is the high sensitivity to detailed aspects of the nonlinear gastrophysics of galaxies, including feedback from short scales to large. It is therefore of interest to try to separate the large-scale coarse-grained patterns that can be predicted from first principle dark matter dynamical methods such as peak patches from the complex response functions (susceptibilities) describing the interior characterizations of haloes, and their embedded galaxies.

In this paper, we focus on the theoretical modelling of the large-scale alignments of haloes. Rather than using general perturbative treatments, we rely on the nonlinear physical model for halo formation exemplified in the peak patch picture, encoding coarse-grained shapes and orientation information in the derived halo catalogues. The detailed cosmic web structure of clusters, filaments, membranes and voids on large scales has been shown to follow by gravitational collapse from the linear fields constrained by the anisotropic linear tides of the most prominent haloes (\cite{bkp96}). First, Section~\ref{sec:PP} gives a general overview of haloes, the peak patch theory of them, and the simulation that yields the halo catalogue used here. In Section~\ref{sec:orientations}, we present calculations of correlations oriented by the shape of haloes using the 3D peak patch simulation and then using 2D projections of this simulation.
Finally, Section~\ref{sec:conclusion} wraps up.
To show how the main features of our orientation-induced findings are quite understandable from an analytic viewpoint, we develop in Appendix~\ref{sec:peaks} an attempt to compute the same correlations from a theoretical point of view using the theory of peaks for Gaussian random fields (BBKS), cast in the peak-patch semi-analytic framework in BM1 which gives a simple but reasonably accurate accounting of the low order statistics derived from the full numerical catalogue.

\section{A statistical description for haloes}
\label{sec:PP}

We first discuss the general issue of cosmological-object catalogues, the difference between halo-finding in final-state space, as done in N-body simulations, and initial-state space, as done in peak-patch simulations. The catalogue statistics are encoded in number density distribution function operators that give masses, initial and final positions, velocities, and of great importance here, the linear tides and their associated strain tensors. Of most interest here are two-point halo-density halo-density averages constrained by strain/tide information which reveal the alignment characteristics of the halo-medium. We then briefly describe the numerical methods we have applied to this study. 

\subsection{Reduced phase space}

A halo catalogue ${\cal C}_{\rm tot} = \cup_c {\cal C}_c $ is a collection of the information on the properties ${\cal C}_c $ of individual cosmic objects $c$.

Let us consider the flow of a bundle of dark matter trajectories from the "initial" linear regime (Lagrangian space) into the "final" nonlinear regime (Eulerian space). If the dark matter starts cold, the 6D phase space is a 3D one, with zero peculiar velocities and it remains restricted to a 3D hypersurface, but trajectories in collapsing regions intertwine, phase wrapping in the 6D phase space. This is seen from the vantage point of a particular Eulerian 3D position in the halo as multiple streams passing through, with a dispersion in velocity as well as a bulk average velocity   {\citep[see e.g][]{2011JCAP...05..015S,2012ApJ...754..126F,2012MNRAS.427...61A}}. From the 3D perspective, the dispersion is seen as heat and haloes are regions of such hot dynamics. Even with the most sophisticated N-body computations it is difficult to simulate the orbits with full accuracy   {\citep[see][for recent attempts at solving the Vlasov-Poisson system]{2016JCoPh.321..644S,2016MNRAS.455.1115H}}. 

In halo-finding, whether in initial or final space, the idea is to identify such regions of hot dynamics, compress the complex information into a few global parameters, such as overall mass, mass-concentration, angular momentum, interior heat and binding energy, and, of special interest to us for anisotropy exploration, tidal tensors, strain tensors and mass inertia tensors.  

The positions ${\bf x}_c$, masses $M_c$ and velocities ${\bf v}_c$ of the haloes $c$ are of course critical to their identification, but there are many other halo-internal measures $Q_{{\rm int},c}^\alpha$ and halo-external measures $E_{{\rm ext}, c}^\beta$ that are useful to characterize in a reduced phase space ${\cal C}_c = \{ M_c, {\bf x}_c, {\bf v}_c,  Q_{{\rm int},c}^\alpha, E_{{\rm ext},c}^\beta \}$. For any single realization a halo catalogue stores these quantities. The entries can all correspond to one redshift, multiple redshifts (e.g., initial linear state and final nonlinearly evolved state), or can label the times by the comoving radial distances of the haloes along the past light-cone, related to their redshifts. Though the sky we observe is one specific past light cone  realization, it is statistically modelled as a member of an ensemble of realizations characterized by a density matrix functional $\rho [u]{\cal D}u$ for all of the degrees of freedom of the system of fields in the cosmic medium $u^A ({\bf x})$, including the fine grained orbit details. The halo catalogue for each specific realization is a set of delta functions of the coarse-grained variables, with the (localized) halo number density operator defined as
\begin{equation}
    \begin{split}
        n_{\rm h,op}({\cal C}) &= \frac{{\rm d}N_{\rm h,op} ({\cal C})}{{\rm d}\ln M {\rm d}^3{\bf x} {\rm d}^3 {\bf v} {\rm d}Q_{\rm int}{\rm d}E_{\rm ext}} \\
         &=\sum_c\delta ({\bf x}-{\bf x}_c) \delta ({\bf v}-{\bf v}_c) \delta (\ln M -\ln M_c ) \\
          &\quad \times \delta (Q_{\rm int}- Q_{{\rm int},c})\delta (E_{\rm ext}- E_{{\rm ext},c}).
    \end{split}
\end{equation}
Typically we have to turn the ${\bf x}$ and $M$ terms into operators acting on the fields, as described below following earlier works such as BBKS for peak positions and BM1 for peak masses. 

Measurements on the data we observe now are at the final times at the collapse redshift of the halo. In the mass-Peak Patch formulation, the measurements are naturally given in terms of initial times in the linear regime. Both the forward mapping from initial to final state and the backward flow from final to initial state can, of course, be done precisely in N-body dark matter simulations since all information is carried along with the particle labels. There is much interest in flowing backward from current data to recover the initial linear state from the final state information, but the orbit entanglements inherent in nonlinear objects limit what is likely to be do-able with just final halo information. Using triaxial ellipsoid collapse in peak patches the forward flow is approximated, though not with enough accuracy to compute, e.g., final state mass-inertia tensors, at least in the current version of the peak patch codes. 

Though one can define haloes in many ways, the most popular in Eulerian space is finding the patches at ${\bf x}_c$ that have their average interior overdensity $M({\bf x}_c, V_{\rm E})/V_{\rm E} $ first upcrossing through a threshold  
$M_c/V_{{\rm E}, c} = \bar{\rho}_{\rm m}\Delta_{M}$, 
as the smoothing volume $V_{E}$ shrinks from large volume to small. Typically one takes $\Delta_{M} =200$. This then defines the halo mass $M_{200, c}$, its volume $V_{{\rm E}, 200, c}$, and hence its radius $R_{{\rm E}, 200, c}$. Some people use the total cosmological mean density $\bar{\rho}_{\rm tot}$, including vacuum energy density (cosmological constant), to define the threshold, with the same $200$ value: that mode of halo identification finds smaller volume and larger mass ones than the ones based on overdensity of the clustering matter relative to the mean density of clustering matter $\bar{\rho}_{\rm m}$. 

Though this is not true for all halo-finding methods, for the standard spherical overdensity method (\cite{BM2}, \cite{Behroozi_2012}), whether determined in final state space or initial state space, as for hierarchical mass-peak patches, the halo positions will be associated with peaks in the mass $\nabla M =0$, and the position operator transforms to 
$\delta ({\bf x}-{\bf x}_c) = \vert {\rm det}(\nabla \nabla M)\vert \delta(\nabla M)$, viewing $M({\bf x})$ as a continuous field. 

Other reasonable criteria for defining haloes exist,
e.g., using binding energy, $\epsilon_{\rm BE}$ in which case the operator would become $\vert {\rm det}(\nabla \nabla \epsilon_{\rm BE})\vert \delta (\nabla \epsilon_{\rm BE})$. This is tightly correlated with mass (BM1), so the haloes found are quite similar. 

For Gaussian filtering the $\nabla M =0$ criterion is the same as requiring that ${\bf x}_c$ is exactly at the centre of mass, but for the top-hat filters that are usually used to measure mass there can be differences.

The haloes are non-Gaussian, so all higher point connected components exist, $\langle \prod_{j=1}^N n_{\rm h,op} ({\cal C}_{c_j} ) \rangle_{\rm cc}$, where, as above,  ${\cal C}_{c_j} $ denotes the collection of reduced phase space variables associated with each halo. Since this paper is focussing on halo anisotropic correlations and alignments, our main targets are  constrained 1-point functions of haloes, which are related to the two-point halo-halo correlation function
\begin{eqnarray}
 \langle  n_{\rm h,op} ({\cal C}_{c_2} )\vert n_{\rm h,op} ({\cal C}_{c_1} ) \rangle &=&  \frac{\langle  n_{\rm h,op} ({\cal C}_{c_2} ) n_{\rm h,op} ({\cal C}_{c_1} )}{\langle   n_{\rm h,op} ({\cal C}_{c_1} ) \rangle}, \\
 1+\xi_{c_2,c_1} ({\bf x}_{c_2 }- {\bf x}_{c_1}) &=&  \frac{\langle  n_{\rm h,op} ({\cal C}_{c_2} )\vert n_{\rm h,op} ({\cal C}_{c_1}) \rangle}{\langle n_{\rm h,op} ({\cal C}_{c_2} ) \rangle},
\end{eqnarray}
relating haloes at position ${\bf x}_{c_2}$ to those at ${\bf x}_{c_1}$. The list of variables may be truncated, 
e.g., with orientation information used in the $c_1$ haloes but marginalized over in the $c_2$ haloes. As well 
the internal variables may be in selected parameter-bands rather than being precisely defined as in the various delta functions. 

The one-point averages $\langle n_{\rm h,op} ({\cal C}_1)\rangle$ can also be used to deal with far field external effects. For example, one of the external variables $E_{\rm ext}$ could be the mass-density field at position ${\bf x}_E$ distant from the halo position ${\bf x}_{c_1}$
\begin{eqnarray}
    \langle E_{\rm ext}({\bf x}_E ) \vert {\cal C}_{c_1} &=&\{M_{c_1},{\bf x}_{c_1},{\bf v}_{c_1}, Q_{{\rm int},c_1}\} \rangle\nonumber  \\
  & =&  \frac{\langle E_{\rm ext}({\bf x}_E) n_{\rm h,op} (\{ {\cal C}_{c_1}, E_{\rm ext}({\bf x}_E) \} )\rangle} {\langle n_{\rm h,op} ({\cal C}_{c_1} ) \rangle}.
\end{eqnarray}
This is the mean field of $E_{\rm ext}({\bf x}_E )$ subject to the constraint of there being a halo with the specified properties ${\cal C}_{c_1}=\{M_{c_1},{\bf x}_{c_1},{\bf v}_{c_1}, Q_{{\rm int},c_1}\}$ at ${\bf x}_{c_1}$. It is related to the usual way of doing correlations,  
$\langle E_{\rm ext}({\bf x}_E) n_{\rm h,op} ({\cal C}_{c_1}) \rangle$, with the averaging involving the joint probabilities at ${\bf x}_E$ and ${\bf x}_{c_1}$. 

\subsection{Halo Tides, Strain and Shear}

If our cluster-object ${\cal C}_{c}$ includes anisotropic tensor information, then the surroundings will have an anisotropic mean field constrained by the cluster's oriented structure, and fluctuations about the mean. We now define the initial linear strains and tides of peak patch haloes which we use to characterize the anisotropy.

Let $\ve{r}$ be the initial comoving position of particles\footnote{  {Note that the Lagrangian position of particles $\ve{r}$ is also often denoted $\ve{q}$ in the literature.}}, $\ve{X}(\ve{r}, t)$ the physical Eulerian position and ${\ve{x}(\ve{r},t)=\ve{X}(\ve{r}, t)/\bar{a}(t)}$  the comoving Eulerian position. The displacement field $\ve{s}(\ve{r}, t)$ at time $t$ is the difference between final and initial positions
\begin{equation}
   \ve{x}(\ve{r}, t) = \ve{r} + \ve{s}(\ve{r}, t).
 \end{equation}
 The associated differential relation giving the response of the infinitesimal Eulerian geodesic deviation $ \delta \ve{x}$ to the stimulus of an initial Lagrangian deviation $\delta \ve{r}$ is $ \delta \ve{x}(\ve{r}, t) = e \delta \ve{r}$, where $e=I+\mathcal{E}$ defines the deformation tensor. It could also be called a response function. 
 
 These relations hold if we smooth the displacement over a coarse-grained initial comoving radius $R_c$ about each point $\ve{r}$:  $\langle \ve{x}\vert \ve{r}, R_c,  t \rangle  = \ve{r} + \langle \ve{s}\vert \ve{r}, R_c, t\rangle $ and  
 \begin{eqnarray}
& \delta \langle \ve{x} \vert \ve{r}_c, R_c, t\rangle = \langle e \vert \ve{r}_c, R_c ,t \rangle  \delta \ve{r}, \\
& \langle \mathcal{E}_j^i \vert \ve{r}_c, R_c, t \rangle = \langle (e_j^i - \delta_j^i )\vert \ve{r}_c, R_c ,t \rangle = \partial \langle \ve{s}^i\vert \ve{r}, R_c, t\rangle / \partial r^j \, .
 \end{eqnarray}
This differential relation of smoothed quantities is also of the stimulus--response form  with the response function $\delta_j^i +\langle \mathcal{E}_j^i  \rangle_c$. The homogeneous  approximation for the finite difference in a sphere uses the local Taylor expansion: 
\begin{equation}
\Delta \ve{X} /\bar{a} = \langle  \ve{x}\vert \ve{r}, R_c \rangle -\langle  \ve{x}_c\vert \ve{r}_c, R_c \rangle  \approx \langle e  \vert \ve{r}_c, R_c , t\rangle ( \ve{r}-\ve{r}_c )\, .
\end{equation}

Thus the (Eulerian) physical deformation of the original Lagrangian sphere coarse-grained over comoving radius $R_{{\rm L},c}$ is $\Delta X^i = \bar{a} \langle e^i_j \rangle_c  R_{{\rm L},c} \hat{R}_{{\rm L},c}^j$, where $\hat{R}_{{\rm L},c}^j$ is a unit vector. This is to be contrasted with the unperturbed Hubble expansion,  $\Delta X^i = \bar{a} \delta^i_j R_{{\rm L},c} \hat{R}_{{\rm L},c}^j$.  Generally $A^i_j = \bar{a}e^i_j$, the local tensorial expansion factor, has a symmetric and antisymmetric tensor decomposition, with the antisymmetric part related to a rotational component. In this paper we are interested in only the symmetric part. It is useful to define the nonlinear strain via a logarithm: $\epsilon_{{\rm NL},c,j}^i= [\ln \langle e \rangle_c]^i_j = \alpha_{{\rm NL},j}^i - \bar{\alpha}\delta^i_j $, where $\bar{\alpha} = \ln a$, encoding the state of strain from an unperturbed expansion reference state,  even in the highly nonlinear regime.  
 
 The linear strain $[\epsilon_{{\rm L},c}]^i_j$ is just the linear version of this deviation. Within the 1LPT (first order Lagrangian perturbation theory) {\it aka} the Zel'dovich approximation,  for initially cold matter the time and spatial components in the displacement field $\ve{s}(\ve{r}, t)$ are separable:
 $$
   \ve{s}(\ve{r}, t) = D(t)\ve{s}(\ve{r}),
 $$
 where $D(t)$ describes the linear growing mode of fluctuations and is normalized to unity at the present time. At the fully linear level in $D$ $ \mathcal{E}^{i}_{{\rm L},j} = \epsilon^{i}_{{\rm L},j}$, but they deviate at 2LPT and higher order. Once nonlinearity becomes strong,  $\epsilon_{\rm NL}$ is a more controlled measure of strain than  $ \mathcal{E}_{\rm NL}$. For example, the "shell crossing" associated with complete collapse along each of the coarse-grained axes is sent off to $-\infty$, where $1+ \mathcal{E}_{\rm NL}$ is zero, and beyond that $1+\mathcal{E}_{\rm NL}$ can go negative. 
 
 The physical velocity $\Delta V$ is related to the physical separation $\Delta X^j$ by a bulk nonlinear Hubble flow and a fluctuation $V_{\rm fluc}^i$ about it: $\Delta V^i =H_{c,j}^i \Delta X^j + V_{\rm fluc}^i$, through the full nonlinear Hubble tensor of the coarse-grained patch $c$,  
 $$
 H_{c,ij} =\partial \Delta \dot{ X}_{c,i} /\partial {\Delta X}_{c,j} = \dot{\alpha}_{{\rm NL},ij} =\bar{H}\delta_{ij}+\dot{\epsilon}_{ij} \, .  
 $$ 
 The Hubble tensor evolves according to 
 $$
 \dot{H}_{ij} +(H^2)_{ij} = -\Phi_{N,ij} -\tau_{{\rm m},ij},
 $$
 where $(H^2)$ is a matrix product, $\Phi_{N,ij}\equiv \partial^2 \Phi_{N} / \partial X^i \partial X^j$ is the gravitational tide and $\tau_{{\rm m},ij}$ is a tide from matter forces, in particular from the pressure tensor. The trace of the gravitational part of the total tide is $4\pi G (\rho_{\rm tot}+3p_{\rm tot})$, where $p_{\rm tot}$ is the pressure, not important for cold dark matter and baryons, but the dominant term for dark energy, and of importance for relativistic and semi-relativistic components.  In the linear perturbation regime the gravitational tide and strain are proportional to each other, 
 $\Phi_{N,{\rm L},ij} = -4\pi G \bar{\rho}_{\rm m} \epsilon_{{\rm L},ij}$. In the nonlinear regime $\Phi_{N,ij}$ has to be self consistently calculated as a function of the instantaneous nonlinear strains, but these are expressible in terms of easily evaluated elliptic integrals. If dark energy is dynamical but not coupled to other matter, it is uniform on subhorizon scales and is otherwise unperturbed, though that unperturbed part does play an important role in how the linear and nonlinear strains evolve, by virtue of its negative pressure slowing the collapse-deformation. 
 
 During cold matter collapse $\tau_{{\rm m},ij}$ is zero, but in an equilibrium final collapsed state with $H_{c,ij} =0$, the two tides balance.  After shell crossing, multiple streams at each Eulerian point $\ve{x}$ lead to the "heat" component of the pressure tensor, $P^{ij}/\rho = \langle  V_{\rm fluc}^i V_{\rm fluc}^j \rangle$  becoming nonzero;   $\tau_{{\rm m},ij}=\partial_i \rho^{-1} \partial_k P^k_j$ (symmetrized in $(ij)$). The variance of the velocity fluctuations about the bulk flow is related to the sound speed $c_{\rm s}^2$. To show this evolution equation is really a familiar one, consider the linear regime with a uniform isotropic sound speed $c_{\rm s} \propto [\langle  V_{\rm fluc}^2 \rangle/3]^{1/2}  $. In that case,  $\tau_{{\rm m}, {\rm L},ij} = c_{\rm s}^2 \nabla^2 \epsilon_{{\rm L}, ij}$, and with $\delta_{\rm L} = -\epsilon_{{\rm L},ii} \propto D(t)$,  $\ddot{\delta}_{\rm L} + 2\bar{H}\dot{\delta}_{\rm L} = 4\pi G \bar{\rho}_{\rm m} \delta_{\rm L} +c_{\rm s}^2 \nabla^2 \delta_{\rm L}$. Hubble drag arises from the linear $H^2$ tensor term, and below the Jeans wavenumber defined by $c_{\rm s}^2 k_{\rm J}^2 = 4\pi G \bar{\rho} $ there is unstable growth, slowed by the Hubble drag, and above there is Hubble-damped oscillation. 
 
 The complexity of shell crossing "heat" generation implies  $\tau_{{\rm m}, {\rm L},ij}$ would need its own equation to develop it, so a drastic model that abruptly turns the gravitational collapse into an equilibrium balance along each principle axis is used in the  BM1 homogeneous ellipsoid approximation for triaxially-collapsing peak patch haloes. Instead of the matter tide being explicitly included, stopping criteria abruptly forcing $H$ to zero with $\alpha_{{\rm NL},ij}$ thereafter frozen are applied to the three axes to arrest the cold flow before $\bar{a}e_{\rm NL}$ passes through zero.  The strain tensor is symmetric, so can be diagonalized. We denote the eigenvalues of $-\epsilon_{{\rm L},ij}$ by $\{\lambda_i\}$, with the principle-axis ordering
 \begin{align}
 \label{order}
\lambda_1 = -\epsilon_{{\rm L},11} \leq \lambda_2 = -\epsilon_{{\rm L},22} \leq  \lambda_3 = - \epsilon_{{\rm L},33}\, .
 \end{align}
The linear overdensity $\delta_{\rm L}({\bf r},R)$ is therefore $\delta_{\rm L}({\bf r},R) = -\Tr(\epsilon_{\rm L})=\lambda_1+\lambda_2+\lambda_3$. The anisotropic part $\epsilon^\prime_{{\rm L}, ij}$ of the tensor and its associated eigenvalues $\{\lambda_i^{\prime}\}$ are defined through  
$  \epsilon_{{\rm L}, ij}= \epsilon_{{\rm L}, ij}^\prime +\delta_{ij}\Tr(\epsilon_{\rm L})/3$.
 There are also eigenvalues for  $-\epsilon_{{\rm NL},ij}=\lambda_{{\rm NL},i}\delta_{ij}$, which are used for the peak patch stopping criteria of BM1 and \cite{2019MNRAS.483.2236S}:  we stop the final 1-axis collapse at a radial freeze-out factor of $f_{\rm coll}=200^{-1/3}\approx0.17$.
 Once that axis compression is reached it is $\alpha_{{\rm NL},11}$ which remains fixed, so the instantaneous nonlinear strain gets progressively more negative since it is relative to the mean cosmological density, which is dropping. Similar stopping criterion for the earlier full collapse of the 3-axis then the 2-axis are used to arrest singularities and shell crossing. The results of 1-axis collapse are encapsulated in a table $\Tr(\epsilon_{\rm L} )({\epsilon_{\rm L}^\prime})$. If we normalize $\epsilon_{\rm L}$ by  $\sigma_0(R_c)$, and define anisotropic  eigenvalue combinations by  $\nu e_v=(\lambda_3 -\lambda_1)/2\sigma_0$ and and $\nu p_v= (\lambda_3 -2\lambda_2 +\lambda_1)/2\sigma_0$, with  $\nu=\delta_{\rm L}/\sigma_0$, the 1-axis collapse criterion translated to initial condition (linear) space takes the form $\delta_{\rm L} = \delta_{{\rm L},{\rm crit}} (e_v, p_v)$.

\subsection{Multi-point Correlation Functions and the Cosmic Web Structure}

\cite{bkp96} described how the main features of the large-scale cosmic web could be understood in terms of mean fields subject to the constraint of the most prominent large-mass haloes oriented by their linear strain tensor (or equivalently their linear tidal tensor).  Multi-point halo constraints (and void constraints) were used, 
\begin{equation}
    \langle E_{ \rm ext}({\bf x}_{E} ) \vert {\cal C}_{c_1}, {\cal C}_{c_2}, ..., {\cal C}_{c_n} ) \, .
\end{equation}
A molecular picture was suggested, by building from 1-halo to 2-halo, showing filamentariness, to 3-halo, showing membranes as well, and to large n-halo constraints, all oriented. As more constraints are added, especially including the oriented information, the more structured $\langle E_{\rm ext}({\bf x}_{E} )\rangle$ becomes, and the smaller the allowed but omnipresent fluctuations $\delta E_{\rm ext}({\bf x}_{E} )$ about the mean field are. In the fully nonlinear Eulerian picture, dealing with multi-point halo constraints is more difficult than for Gaussian random fields. The mean field plus allowed fluctuations subject to multiple prominent mass-peak constraints has been useful in setting up N-body and gas-dynamical simulations, e.g., \cite{wb97}. 

In the linear regime of small $E_{\rm ext}({\bf x}_{E})$ the mean can be written in terms of linear response functions $\beta_{E,c_j,\alpha}$
\begin{multline}
 \langle E_{\rm ext}({\bf x}_{E} ) \vert \cup {\cal C}_{c_j}) \rangle 
 \rightarrow \sum_{j}\int {\rm d}^3{\bf x}_{c_j}{\rm d }Q^\alpha_{c_j}P_{\rm sel}(Q^\alpha_{c_j})\\
\times \beta_{E,c_j,\alpha} ({\bf x}_{E}-{\bf x}_{c_j}) Q^\alpha_{c_j} n_{c_j}({\bf x}_{c_j},Q^\alpha_{c_j}) 
\, ,
\end{multline}
where $P_{\rm sel}(c_j\vert Q^\alpha_{c_j})$ is a selection function acting on the internal parameters. The response functions 
\begin{equation}
\beta_{E,c_j,\alpha} = \frac{\delta E_{\rm ext}({\bf x}_{E})}{\delta Q^\alpha_{c_j} n_{c_j}({\bf x}_{c_j} )}\, , \nonumber
\end{equation}
are defined in terms of functional derivatives. Their inverses are often referred to as bias functions, and the expansion in terms of responses acting on halo number densities is a generalized form of the halo-model.  Added to this mean field is a fluctuating residual, which exists both outside and inside of the haloes. 

This extended use of the regular halo one-point function describes how continuous fields $E_{\rm ext}({\bf x}_{c_2} )$ at ${\bf x}_{c_2}$ are influenced by a halo at ${\bf x}_{c_1}$. Of course we do not have to carry the $E_{\rm ext}({\bf x}_{E})$ variable information in the distribution function to get this result, just cross-correlate $E_{\rm ext}$ with the cluster-density operator, $n_{\rm h,op} $. Once we go to halo-halo correlations, the non-local complexity of exclusion means we would not want to carry $n_{\rm h,op} ({\bf x}_{c_2})$ information in an extended one point halo distribution function centered at ${\bf x}_{c_1}$.  Instead to treat halo auto-clustering we do direct halo-halo two-point functions, and, if needed,  higher connected $N$-point functions.   

\subsection{The mass-Peak Patch method}

$N$-body simulations often require demanding computational time to get a series of light-cone snapshots of the formation of large scale structures across cosmic time in a large cosmological volume with enough mass and/or spatial resolution. In a series of papers in the late 80s and early 90s, Bond and Myers synthesized the random field theory of peaks (BBKS) and excursion sets \citep{BCEK} into the hierarchical mass-peak picture, with algorithms and computational methods given in \cite{ref1} (BM1), validation in \cite{BM2} and applications to the thermal Sunyaev-Zeldovich effect and X-ray emission in \cite{BM3}, and to the cosmic infrared background in \citep{1993egte.conf...21B}. The method has continued to be used ever since, for hydrodynamical and N-body simulations of fields constrained by the mass-peak information, map-making for all sorts of applications. In particular the rise of huge datasets from large scale structure and cosmic background experiments has stimulated simulations of light cones on massive scales, and the associated construction of {\it webskys}, fully correlated  maps in 2D and 3D (redshift) space for a large variety of signals. The further developments  of the algorithm and its validation in this new era are described in \cite{2019MNRAS.483.2236S} and the websky applications in \cite{SABx2}.

The conceptual picture is one of haloes as regions of hot dynamics that are bundled. 
Each light cone constructed using the hierarchical mass-Peak Patch method contains the large scale superclustering. 
Our target here is to 
extract from 
the initial distribution of matter (Lagrangian space) the regions that are likely to collapse and form a halo in the final state (Eulerian space). Here, one can decouple the very nonlinear gravitational collapse of matter from the large-scale flows. Hence, starting from a linear density field, the mass-Peak Patch algorithm can be split into three stages:
\begin{enumerate}
 \item Peak patches: from the initial distribution of matter, we look for non-nested regions\footnote{Here, non-nested collapsing regions refer to the largest scale which can collapse according to the ellipsoidal model, therefore rejecting cloud-in-cloud effects (substructures).} that are likely to collapse and form a virialized object according to an ellipsoidal collapse dynamics, \textit{ie} the peak patches. In practice, this calculation is only performed at peaks of the density field on a hierarchy of scales to reduce computational demands, as this has been found to be sufficient to find the complete list of non-nested regions for most use-cases.
 \item Exclusion: from the list of tentative peak patches, binary exclusion is then implemented to ensure non-overlapping patches to avoid double counting mass.
 \item Displacements: finally, displacements of these peak patches averaged over the Lagrangian volume of the halo are computed using the Zel'dovich approximation or higher orders (2LPT) when needed.
\end{enumerate}
  {We refer to the header of Section~2 of \cite{2019MNRAS.483.2236S} for a short summary of the key aspects of the peak patch formalism. The reader will also find in the rest of the same article a detailed presentation of the algorithm as well as an extensive statistical validation of the halo catalogues compared to full N-body results.}

\subsection {Characteristics of the simulation}

\begin{figure*}
	\centering
	\includegraphics[width=0.92\columnwidth]{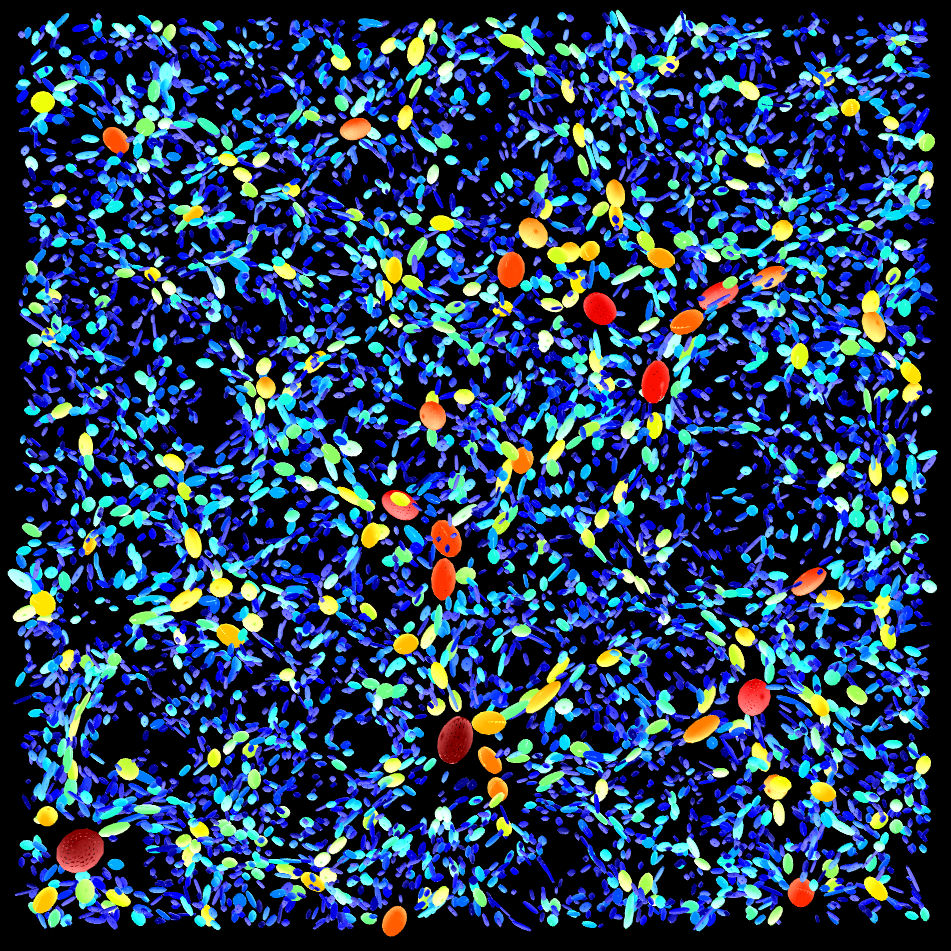}
		\includegraphics[width=0.92\columnwidth]{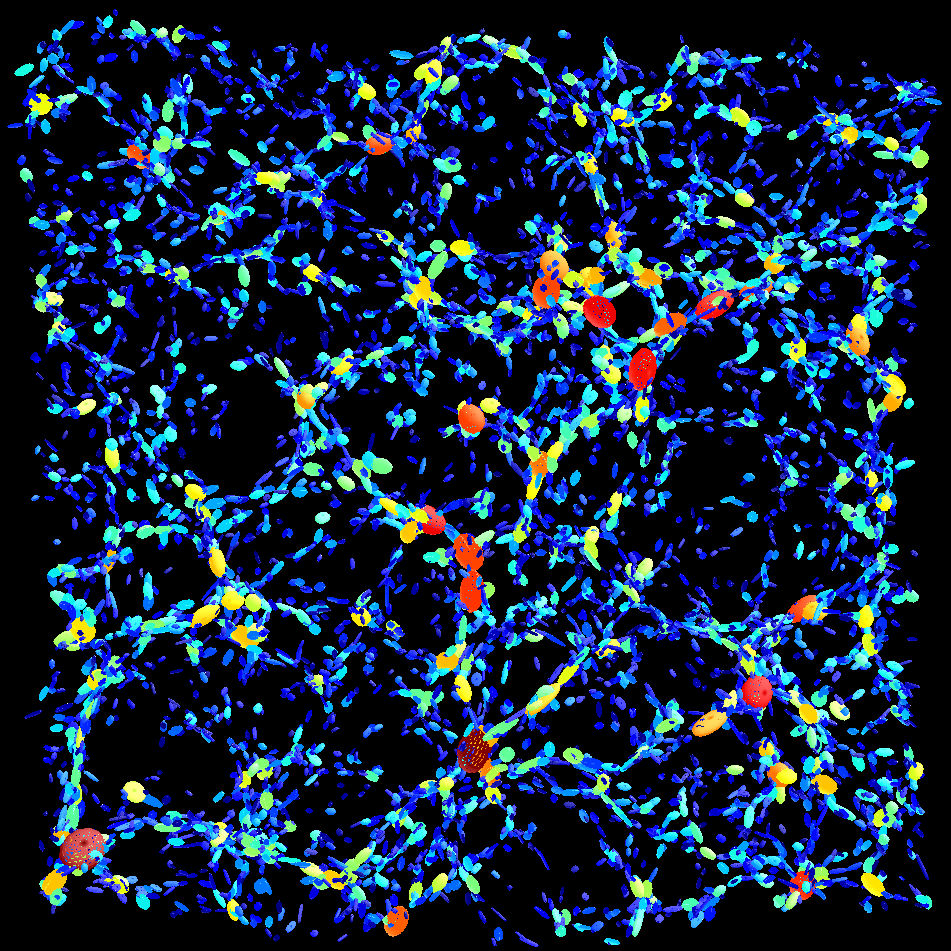}
		\includegraphics[width=0.19\columnwidth]{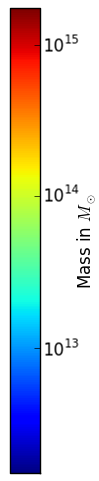}
	\caption{Visualisation of haloes from the peak patch catalogue as projected ellipsoids based on the anisotropic strain tensor in the corresponding Lagrangian patches. Left-hand panel: haloes from a $512\times512\times25~\text{Mpc}^3$ slice in Lagrangian space. Right-hand panel: same haloes moved to Eulerian space. Colours encode the mass as labelled, even though the volume of the ellipsoids is also proportional to the mass.
	}
	\label{strain-vis}
\end{figure*}

In the peak patch simulation of dark matter used here, haloes are identified within a large comoving periodic box at a specified redshift. For this paper we ran a $(3000~\text{Mpc})^3$ simulation using $4096^3$ initial particles and output halo catalogues at $z = 0$. The final catalogue contains more than 32 million haloes above a minimum halo mass of 100 simulation cells, or $1.5\times 10^{12}~M_\odot$. This run was not optimized for performance or memory requirements, but had a run time of 1.0 hour on 770 Intel ``Skylak'' 2.4 GHz cores of SciNet's Niagara cluster \citep{niagara}, for a total runtime of 700 hours and a peak memory footprint of 3.5 TB. This is over a thousand-fold speedup relative to $N$-body simulations of the same size.  Our $\Lambda$CDM cosmological parameters are compatible with Planck 2018 results \citep{planck2016-l01}: $\Omega_\mathrm{m} = 0.31$; $\Omega_\mathrm{b} = 0.049$; $\sigma_8 = 0.81$; $n_\mathrm{s} = 0.965$; $h = 0.68$. 

For the purposes of this work, a peak patch simulation gives, for each halo, a Lagrangian position and an Eulerian position (determined by its 2LPT displacement), a Lagrangian radius $R_\mathrm{L}$, and a strain tensor $\Epsilon$ computed in Lagrangian space. The mass of a halo is determined from its Lagrangian radius through the mean density of the Universe, 
\begin{equation}
\label{eq:halomasses}
    M_\mathrm{h} = \frac{4}{3}\pi R_\mathrm{L}^3\Omega_{\rm m}\rho_{\mathrm{c}, 0}\ \text{ with }\rho_{\mathrm{c}, 0} = \frac{3H_0^2}{8\pi G}.
\end{equation}
Throughout this work we consider three subsets of the halo catalogue which simply correspond to three different mass selections:
\begin{enumerate}
    \item low masses haloes with mass $M\in [1.5\times 10^{12} M_\odot, 10^{13}M_\odot)$;
    \item intermediate masses haloes corresponding to $M\in [10^{13}M_\odot, 10^{14}M_\odot)$;
    \item high masses haloes with $M \geq 10^{14}M_\odot$.
\end{enumerate}

\section{Orientation of haloes in the peak patch picture}
\label{sec:orientations}

Since the cosmic web develops from linear tidal anisotropies in the initial Gaussian random field which are correlated across scales, initial strains at halo positions are very likely to be strongly aligned with the large-scale cosmic web. This is what we  investigate in detail in this paper.

Even though the initial strain tensor is not equivalent to the actual shape of the dark matter haloes at the present day, a number of theoretical models are based on this quantity (e.g., linear tidal alignment model, tidal torquing or any generalisation of these two based on perturbation theory). Because the strain and tide are proportional, and strain is what defines coarse-grained collapse on halo-scales, we think it is the most important indicator for collapse. By contrast the initial moment of inertia tensor, while highly correlated with the tide in linear theory, is generally mismatched because of the $r^2$ weighting of the strain compared to the unweighted strain average that we want to concentrate on. This mismatch is important since it is acted upon by the tides in the perturbative treatment of angular momentum generation by tidal torques, which is quadratic in linear amplitudes. 
Given this picture spatial correlations among the initial strain tensors should therefore induce correlations between the (observable) late-time halo or host galaxy shapes. 
Modelling this initial large-scale tidal coherence is the first building block in any model of intrinsic alignments and is the purpose of this article  {, notably motivated by the fact that initial conditions have been shown to determine very accurately the late-time shape and spin of dark matter haloes \citep[see e.g][for a recent study]{2020arXiv201202201C}. }
From there one can then try to model the relationship between the initial strain and the late-time morphology and orientation of galaxies. This is however a complex and very challenging issue as it involves nonlinear physics, including for instance highly anisotropic accretion together with a precise understanding of the baryonic physics engaged in the patch evolution,  and, alas,  is  not within reach of first-principle calculations. Cosmological hydrodynamical simulations that resolve the details of galaxy physics together with their connection to the large-scale environment may be needed and is at the core of many developments in the recent years. This part of the modelling is thus for the future.

In this section, we first visualise the shape of peak patch haloes as given by their initial linear strain tensor -- denoted in this work by $\Epsilon$ or the related $\epsilon = \ln (I+\Epsilon)$, $\approx\Epsilon$ to linear order -- before moving to more quantitative measurements.

\subsection{Visualisation of haloes intrinsic alignment}
\label{visualisation}

\begin{figure}
	\centering
	\includegraphics[width=\columnwidth]{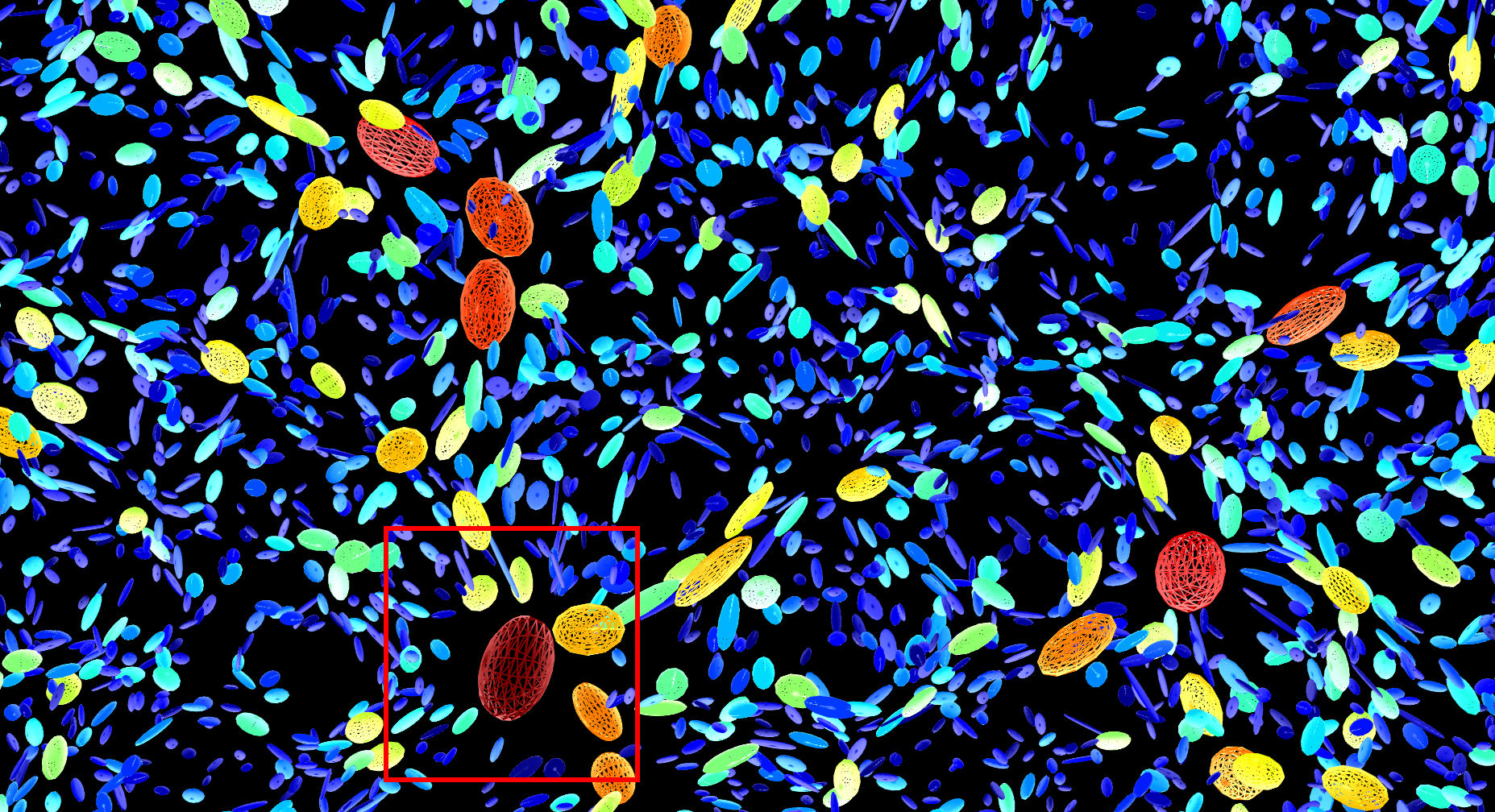}
	\vskip 0.3 cm
		\includegraphics[width=\columnwidth]{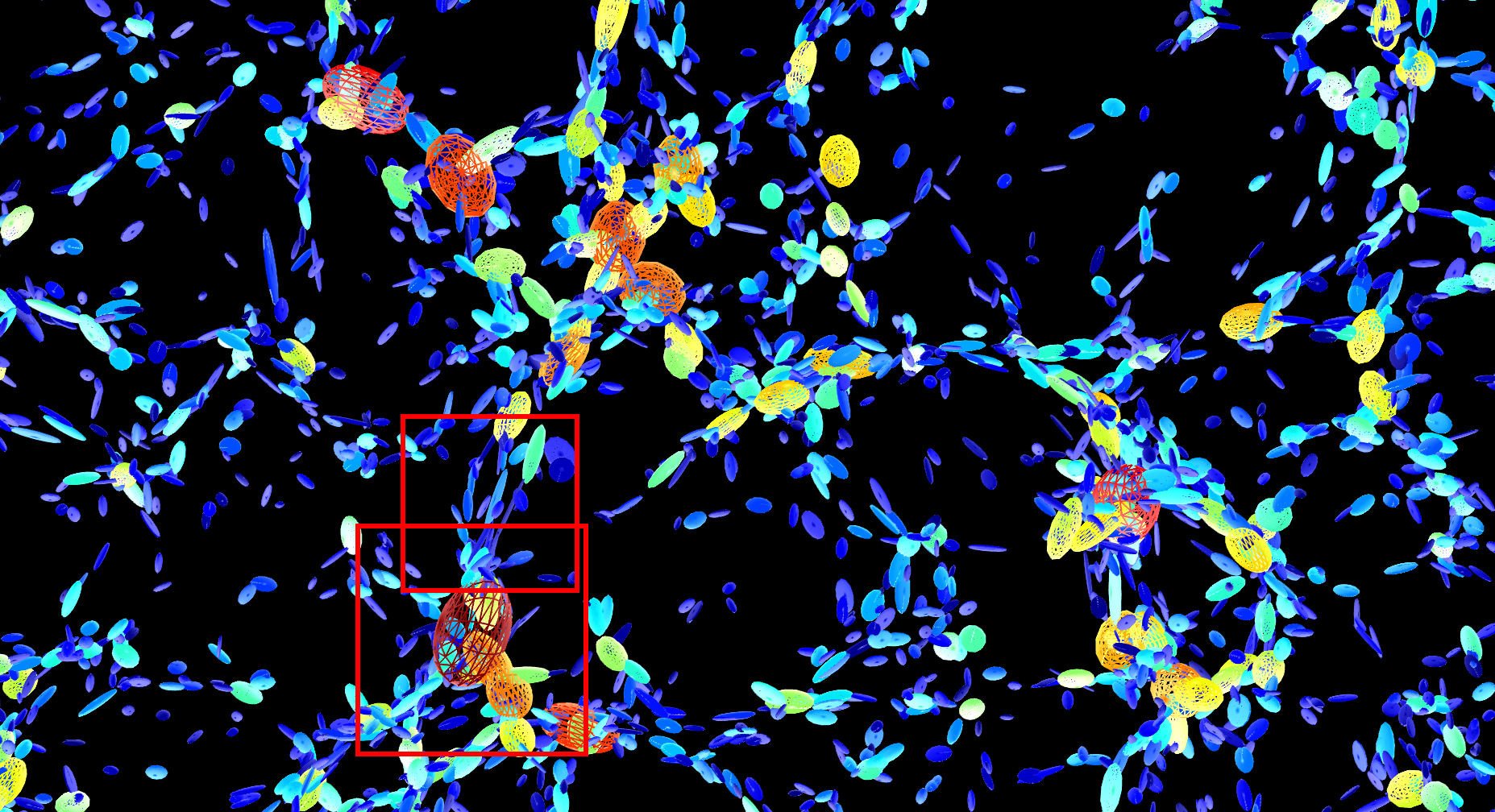}
		\caption{Top panel: Zoom over the same slice of as Fig.~\ref{strain-vis} in Lagrangian space. Bottom panel: Same as top panel but in Eulerian space. Red squares mark regions from which videos, that are introduced in the main text, show 3D visualizations.
			\label{strain-vis2}}
\end{figure}

Fig.~\ref{strain-vis} shows haloes in a $25~\text{Mpc}$ thick slice of the simulation, both in Lagrangian space (left-hand panel) that is to say in initial condition space, and in final state (Eulerian) space (right-hand panel) once an adaptive (cluster-specific) 2LPT displacement of the haloes is performed\footnote{This visualisation made use of the Mayavi 3D Python visualisation library~\citep{ramachandran2011mayavi}.}. 

 For visualization we represent a halo by an ellipsoid whose principal axes follow the ones of the initial anisotropic strain tensor $\Epsilon$ (or, equivalently tidal tensor). Their lengths are given by
\begin{equation}
\label{eq:ellips}
  C_i = \frac 1 2\exp\left(- \lambda_i'\right)R_\mathrm{L},
\end{equation}
so as to keep the volume proportional to the mass and to capture the deforming effects of Zeldovich dynamics \citep{1970A&A.....5...84Z} at first order while avoiding singularities.
In what follows, we will call the major direction of the halo the one with the longest semi-axis $C_i$ and the minor direction one with the smallest semi-axis. As we have ordered the eigenvalues of the strain tensor (see equation~(\ref{order})), the major direction is then referring to the direction of the smallest eigenvalue of the strain tensor $\lambda_1$ such that
\begin{equation}
  C_3 \leq C_2 \leq C_1.
\end{equation}
This choice is not \textit{ad hoc} since indeed haloes and galaxies are believed to undergo tidal stretching by the cosmic web such that their late-time shape is indeed tightly connected to the shear they experienced early on.
The colour coding used in this figure is an indicator of the mass of the haloes, from blue (low mass) to red (high mass) as labelled, which is directly related to the radius (see Eq.~(\ref{eq:halomasses})). Note that the volume enclosed in the ellipsoid is proportional to $R_{\rm L}^3$ and therefore to the mass. Hence, bigger haloes are the red ones (the more massive objects) and smaller haloes are the blue ones.

\begin{figure*}
	\centering
	\includegraphics[width=0.9\textwidth]{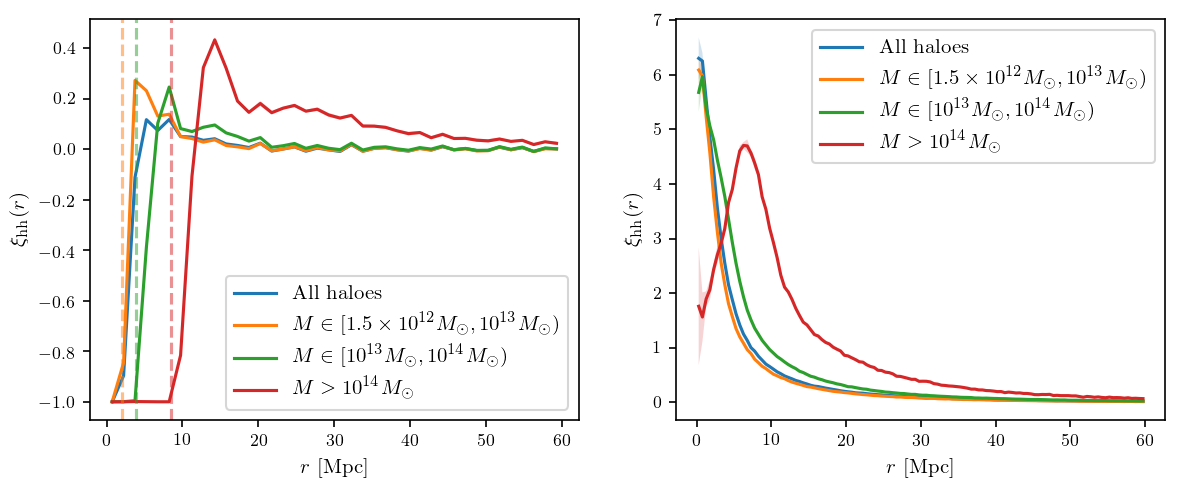}
	\caption{Two-point density-density isotropic correlations in Lagrangian space (left-hand panel) and Eulerian space (right-hand panel) for different bins of mass as labelled. Vertical dashed lines show minimal $R_\mathrm{L}$ values for each bin of mass. The Eulerian analogue involves compression to an overdensity of $200$, so these lines would be multiplied by $\sim (200)^{-1/3} \approx 0.17$, significant in this plot only for the largest mass bin where exclusion effects are still evident even after the dynamics draws the haloes together.}
	\label{2Iso}
\end{figure*}

The emergence of the cosmic web is clearly seen on this figure, with large voids almost empty of haloes surrounded by walls and filaments which connect onto the nodes of the cosmic web.
As expected, the most massive objects lie in the dense nodes of the web while smaller structures tend to reside mostly within filaments \citep[e.g][among many others]{1974ApJ...194....1O,1984ApJ...284L...9K,1989MNRAS.237.1127C,ATTT}.
The representation of halo shapes shows remarkably well-aligned haloes with their surrounding denser filaments and walls. This alignment is subsequently enhanced by the collapse of structures (right-hand panel) but is already in place in the initial conditions.
Even in less dense regions, the major direction of haloes seems to be strongly correlated amongst neighbouring haloes.

Fig.~\ref{strain-vis2} displays a zoom onto a filament, again in Lagrangian (top panel) and Eulerian space (bottom panel).
In addition to the 2D projections shown in the paper, we provide 3D visualisations online so as to rotate the scene and get a more accurate view of the alignments of structures. 
In particular, movies rotating around a halo in Lagrangian space and Eulerian space together with a filament in Eulerian space
-- as delineated with the red squares in Fig.~\ref{strain-vis2} -- 
can be respectively accessed
at \url{https://youtu.be/1UOH8jaaQYU}, \url{https://youtu.be/dD4_vXHIk6o} and \url{https://youtu.be/NodNFozaig8}.
The filaments of the cosmic web seem again to be well-aligned with the directions of the initial anisotropic strain tensor and therefore potentially with halo and galaxy shapes at low redshift.

In the remainder of this article, we will quantify these intrinsic alignments as predicted by the peak patch model. We will first focus on the standard isotropic halo clustering before introducing halo anisotropic clustering in the frame of the cosmic web by means of the frame of the initial strain and eventually investigate initial shape alignments.

\subsection{Isotropic clustering of haloes}

Before studying the orientation of haloes, let us first quantify the isotropic clustering of peak patch haloes by means of the two-point density-density isotropic correlation function defined as
\begin{equation}
  \xi_{\mathrm{hh}}(r) = \langle \delta_\mathrm{h}(\ve{r^{\prime}}) \delta_\mathrm{h}(\ve{r^{\prime}} + \ve{r})\rangle = \frac{\langle n_\mathrm{h}(\ve{r^{\prime}}) n_\mathrm{h}(\ve{r^{\prime}} + \ve{r})\rangle}{\langle n_\mathrm{h} \rangle^2} -1,
\end{equation}
where $\delta_{\rm h}$ is the number density contrast of haloes, \textit{i.e.} $\delta_\mathrm{h}(\ve{r}) = {n_\mathrm{h} (\ve{r}) }/{\langle n_\mathrm{h} \rangle}-1$, and $n_\mathrm{h}$ is the number density of haloes. Note that the halo correlation function $\xi_\mathrm{hh}$ only depends on the modulus of the pair separation $\ve{r}$ if we assume statistical homogeneity and isotropy of the random fields. By definition, this function can range from -1 (anti-correlation/exclusion) to $+\infty$ (high correlation).

For our catalogue, we stack halo pairs depending on their pair separation using regular bins in $r$ (we call $\Delta r$ the corresponding radial step size). We then estimate $\xi_\mathrm{hh}$ by comparing this stacking to the total number of halo pairs if the haloes were uniformly distributed in the simulation box, weighted by the volume fraction of the bins. To estimate measurement uncertainty, we divide the simulation volume in 8 and compute separate estimations of $\xi_\mathrm{hh}$ for each of these sub-volumes. The final estimation of $\xi_\mathrm{hh}$ is the mean of the 8 separate estimations, and the corresponding uncertainties are computed as the standard error on the mean.

The resulting halo correlation function is shown in Fig.~\ref{2Iso} for different halo masses, namely
i) low masses with $M \in [ 3.5 \times 10^{13} M_{\odot}, 10^{14} M_{\odot}]$ (orange),
ii) intermediate masses with $M \in [10^{14} M_{\odot}, 10^{15} M_{\odot}]$ (green),
and iii) high masses with $M > 10^{15} M_{\odot}$ (red).
The left-hand panel shows the measurements in Lagrangian space (with $\Delta r = 1.5~\mathrm{Mpc}$) while the right-hand panel displays the counterpart in Eulerian space (with $\Delta r = 0.5~\mathrm{Mpc}$). The vertical dashed lines show the minimal $R_\mathrm{L}$ values for each bin of mass. Note that the radial step size of these measurements is higher for Lagrangian measurements to prevent pixelization effects due to the limited resolution of the simulation.

In Lagrangian space we see a clear exclusion zone -- when the correlation function goes to -1 -- that is larger for massive haloes, depending upon a combination of the cluster's Lagrangian radius $R_{{\rm L},c}$, large for larger masses, and an average of the radii $R_{{\rm L},c^\prime}< R_{{\rm L},c}$ of all other lower-mass haloes.  
Peak patches explore three types of exclusion, half-exclusion allowing for clusters anywhere outside of $R_{{\rm L},c}$, full exclusion enforcing a larger disallowed zone, of radius $R_{{\rm L},c^\prime} +R_{{\rm L},c}$, and binary exclusion that is a compromise between the two: clusters in the in-between region have a mass-conserving apportionment of space to the two clusters in question, but this is done pairwise, whereas simultaneous apportioning of space of all clusters in the in-between region would be better, albeit statistically complex. It is difficult to do the binary exclusion adopted for our halo catalogue results, and, indeed, for full exclusion, in any semi-analytic framework, so there is little recourse but to do numerical estimates of the clustering such as those reported here to be quantitative about exclusion. As we vary the target masses of the haloes, $R_{{\rm L},c}$ ranges from a few megaparsecs at the lowest masses we consider to about 10 Mpc for the most massive clusters. Exclusion is a very nonlinear effect, involving in general full clustering information locally of the clusters, to ensure haloes are non-overlapping in Lagrangian space, a hard sphere exclusion. On top of that, the statistics of the initial field typically prevent the formation of deep enough potential wells too close to each other, simply because of topological arguments \citep{2016MNRAS.456.3985B,2018MNRAS.479..973C}.
Beyond the exclusion zone the correlation function displays a positive enhancement, a biasing bump,  before it converges towards zero on tens of megaparsecs scales, following a perturbative bias model  \citep{2019arXiv191009561M}. As is evident, the linear bias, which measures the linear response of the halo density field to a mass density perturbation,  is larger for larger mass haloes \citep{1984ApJ...284L...9K,2018PhR...733....1D}.

Once haloes are displaced according to second order Lagrangian perturbation theory, the exclusion zone is progressively filled as haloes move and large structures collapse. Only the most massive haloes still display an exclusion zone larger than the resolution of our simulation at redshift zero. On large scales, gravitational collapse enhances halo clustering as expected, with more massive objects having a larger bias.

\subsection {Halo clustering in the strain eigenframe}

One of the obvious drawbacks of a standard isotropic halo correlation function is that it mixes all the directions of the cosmic web so that the quite anisotropic clustering, e.g.,  along filaments,   is averaged out. To highlight the clustering anisotropy, one can mark the two-point halo-halo correlation function by adding  constraints on the pair members. In this section, we use knowledge of the target cluster's strain tensor to compute the mean number density profile of haloes around it, for now not taking into account orientation aspects of the non-target clusters that we stack on, although we do that later. In this work the shapes of haloes are modelled as ellipsoids, with axes deformed from spherical by the time-dependent nonlinear strain tensor, but the late-time deformations are determined completely by the pre-collapse linear initial strain. That is, we use the target's strain frame to define directions in the local cosmic web and compute the {\it strain-tensor-oriented correlation function} within this frame. Hence we can investigate alignment of cluster-scale tides with the cosmic web, expecting to see differential clustering along filaments of the haloes rather than perpendicular to them. 

\subsubsection {{\it }Strain-tensor-oriented correlation functions via oriented stacking}
\begin{figure*}
	\centering
		\includegraphics[width=\columnwidth]{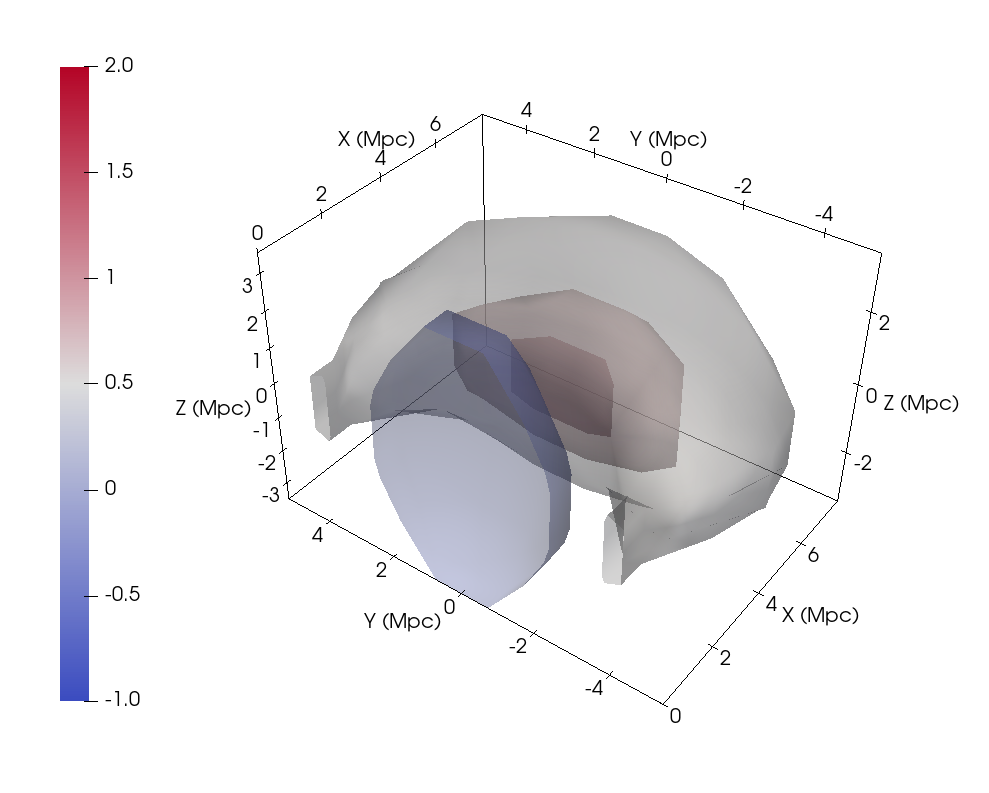}
			\includegraphics[width=\columnwidth]{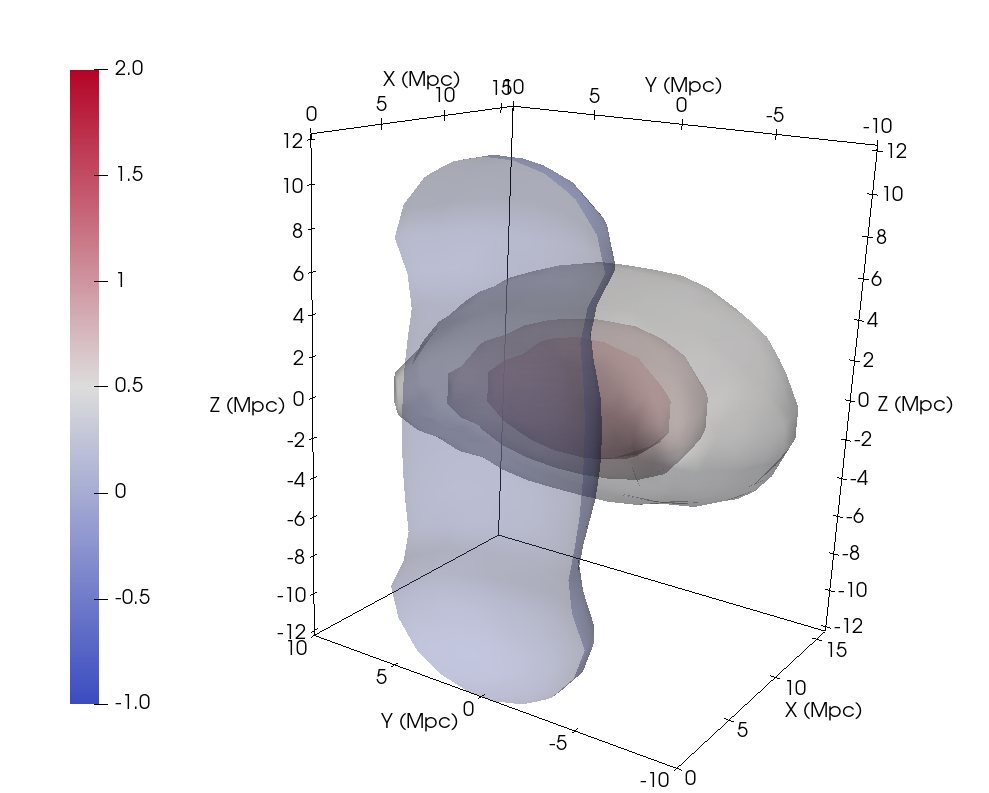}\\
			\includegraphics[width=\columnwidth]{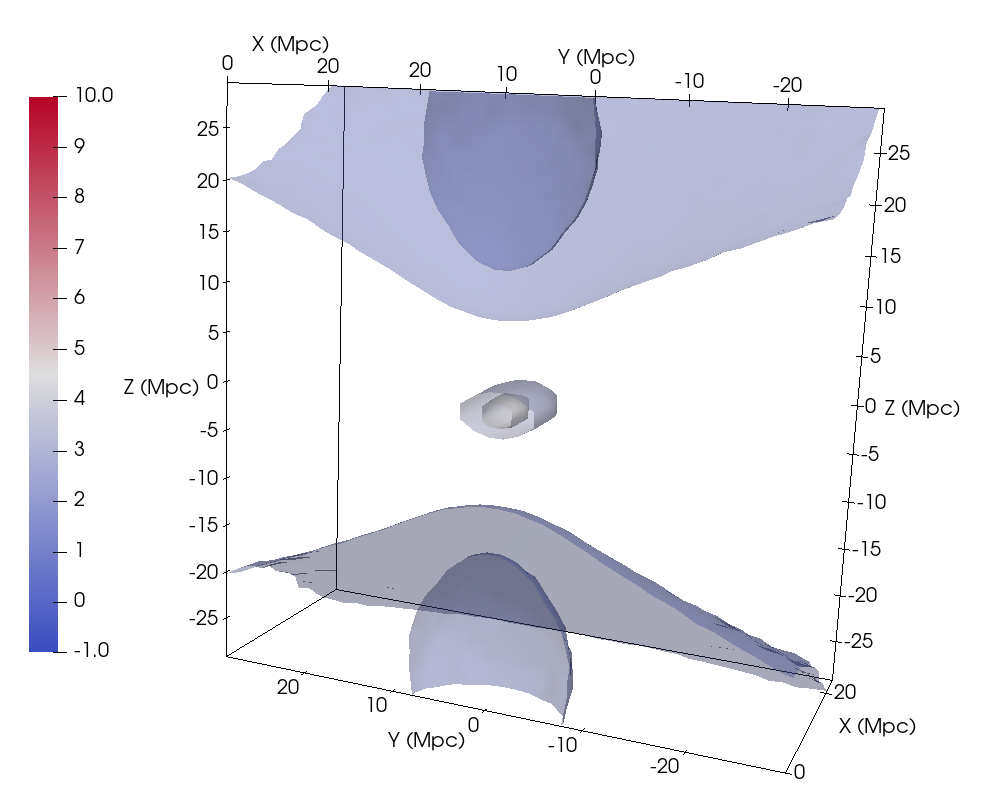}
				\includegraphics[width=\columnwidth]{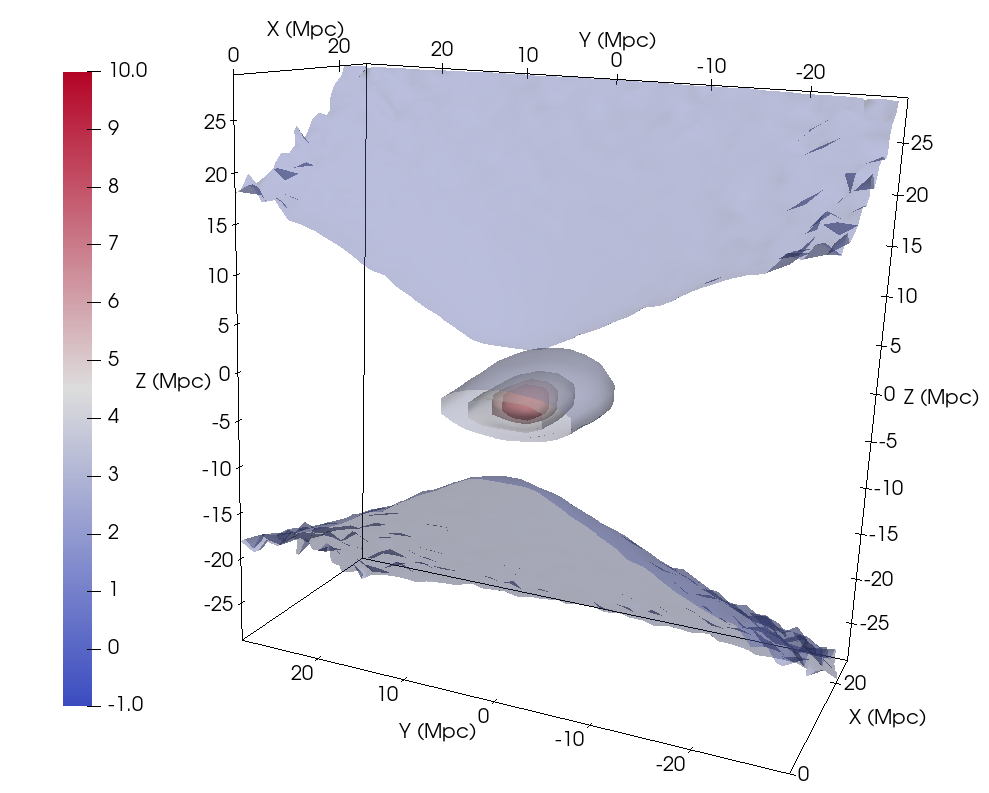}
	\caption{Stacking of haloes on strain-oriented haloes yields the $\xi_\mathrm{hh}^{\Epsilon}(\ve{r})$ correlation function, shown here with a voxel size $\Delta x = 1.5$~Mpc.
	The top panel shows the measurements in Lagrangian space while the bottom shows Eulerian space measurements. Two mass bins are displayed: $1.5 \times 10^{12}$--$10^{13}M_{\odot}$(left-hand panels) and $10^{13}$--$10^{14}M_{\odot}$ (right-hand panels). Here, the $X, Y, Z $ axes respectively refer to the major, intermediate and minor directions of the selected haloes. We show only a few isocontours for the sake of clarity. In Lagrangian (respectively Eulerian) space, we show contours for $\xi_\mathrm{hh}^{\Epsilon} = -0.5, 0.4, 0.8, 1.2$ (resp. $\xi_\mathrm{hh}^{\Epsilon} = -0.1, 0, 2, 4, 6, 8$).}
	\label{3D-DD}
\end{figure*}

\begin{figure*}
	\centering
    \includegraphics[width=\textwidth]{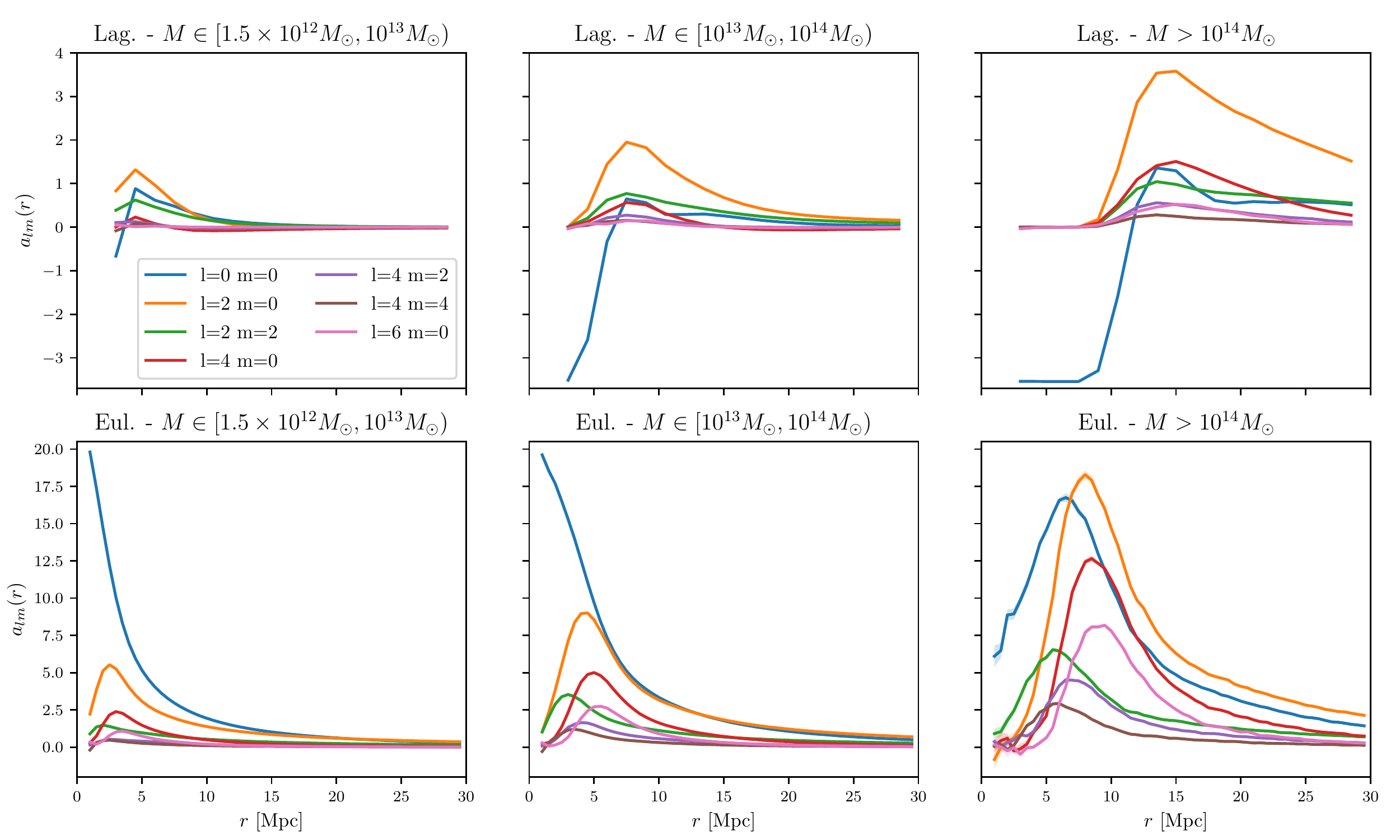}
	\caption{$Y_l^m$ decomposition of the strain tensor oriented halo correlation function $\xi_\mathrm{hh}^{\Epsilon}(\ve{r})$ in shells of various radii. Results in Lagrangian space (top panels) and Eulerian space (bottom panels) are displayed for different mass bins from left to right as labelled. The legend gives the corresponding $(l,m)$ values.}
	\label{ylmDec}
\end{figure*}

To compute the two-point strain-tensor-oriented halo-halo correlation function $\xi_\mathrm{hh}^{\Epsilon}$ we use the natural basis for each selected halo defined by the principal directions of the strain tensor $\Epsilon$, as explained in Section \ref{visualisation}, and we stack its neighbouring haloes in this frame in a procedure we call {\it oriented stacking}, the goal being to quantify anisotropic behavior following the eigendirections of the target haloes. The correlation between two haloes at $\ve{r}_{h1}$ and $\ve{r}_{h2}$ only depends upon $ \ve{r} = \ve{r}_{h2} -\ve{r}_{h1}$:
\begin{equation}
 \xi_\mathrm{h1,h2}^{\Epsilon,h1}(\ve{r}) = \langle \delta_\mathrm{h} \left(\ve{r}_{h1}\right) \delta_\mathrm{h} \left( \ve{r}_{h1} + {\cal R}_{\Epsilon,h1} \ve{r})\right)\rangle,
\end{equation}
where ${\cal R}_{\Epsilon,h1}$ is the orthogonal matrix that diagonalizes $\Epsilon_{h1}$, the strain tensor of a halo at position $\ve{r}_{h1}$, and therefore represents the change of basis from the frame of the simulation to the eigenframe of the strain tensor, such that
\begin{equation}
  -\Epsilon = {\cal R}_{\Epsilon}
  \begin{pmatrix}
	\lambda_1 & 0 & 0 \\
	0 & \lambda_2 & 0 \\
	0 & 0 & \lambda_3
  \end{pmatrix}
  {\cal R}_{\Epsilon}^T \text {\hspace{0.2cm} with \hspace{0.2cm}} \lambda_1 \leq \lambda_2 \leq \lambda_3.
\end{equation}
For the sake of simplicity, we will use the shorter notation $\xi_{h h}^{\Epsilon}(\ve{r})\equiv \xi_\mathrm{h1,h2}^{\Epsilon,h1}(\ve{r})$ in the remainder of this paper. The difference with the previous isotropic correlation is that we are now deliberately breaking the isotropy by enforcing the halo eigenframe information upon the correlation statistics.
Hence we now do the stacking in 3D instead of in 1D as was done with the isotropic case before when the correlation only depended upon the modulus of the pair separation. For the Lagrangian space results we adopt a regular grid with voxel size $\Delta x = 1.5$ Mpc to get smoother results. For Eulerian space $\Delta x$ can be smaller. 

\subsubsection{3D oriented clustering}

Fig.~\ref{3D-DD} shows a 3D visualisation of the oriented halo correlation function for two bins of mass. The X, Y and Z axes respectively refer to the major, intermediate and minor principal directions of the strain tensor.
Because  of  the  arbitrary  choice  of  orientation  of  the eigenvectors, the 3D oriented correlation function must be invariant under any rotation of $\pi$ around these principal directions. The top panels of Fig.~\ref{3D-DD} show the 3D oriented correlation function in Lagrangian space. The contours in blue show the exclusion zone near the central halo. This exclusion is very anisotropic and is stretched along the minor axis of the halo. The clustering that is maximum in the red region is also very anisotropic and occurs along the major direction. This is a clear signature of the cosmic web like environment with typically a filament along the major direction where most neighbouring haloes reside, a wall that encompasses the filament and tend to lie in the plane formed by the major and intermediate direction while two voids tend to be located in the minor eigendirection on both sides of the wall.
The bottom panels show the oriented correlation functions in Eulerian space. The exclusion zone has been filled by the displacement of haloes and the anisotropy of the clustering is enhanced with large voids growing on both sides of the wall and the filament along the major axis being even more pronounced and sharper as a result of the gravitational collapse of structures.

This corroborates what we were expecting from the halo shape visualisations of Fig.~\ref{strain-vis}. The major direction of haloes more likely points towards other haloes while their minor direction points towards neighbouring voids. Let us also note that the bump in density is approximately at the same distance, around 5-10 Mpc, found for the isotropic correlation (Fig.~\ref{2Iso}), quantifying the typical extent of inter-cluster filaments (which is an increasing function of halo mass). 
  {Note that this anisotropic clustering of haloes was also investigated recently in the context of galaxy-galaxy lensing \citep{Osato_2018} where similar results were found.}

\subsubsection{Harmonic analysis of the oriented clustering}

Since 3D correlation functions have obvious difficulties for visualization,  we will now project the 3D correlation maps via a harmonic decomposition: for each successive spherical cut at fixed radius of $\xi_\mathrm{hh}^{\Epsilon}$ field  we expand $\xi_\mathrm{hh}^{\Epsilon}$ in the spherical harmonics basis $\{Y_l^m\}$: 
\begin{equation}
  \xi_\mathrm{hh}^{\Epsilon}(r, \theta, \varphi) = \sum_{l=0}^\infty \sum_{m=-l}^l a_{l,m}(r) Y^m_ l(\theta, \varphi) \, , 
\end{equation}
with radially dependent amplitudes
\begin{equation}
  a_{l,m}(r) = \int_{\varphi=0}^{2 \pi} \int_{\theta=0}^\pi \xi_\mathrm{hh}^{\Epsilon}(r, \theta, \varphi) \overline{Y_l^m(\theta, \varphi)} \sin (\theta) \; d \theta \; d \varphi,
\end{equation}
The $\{Y_l^m\}$ convention we adopt, in terms of associated Legendre polynomials $P^m_l$, is 
\begin{equation}
  Y^m_l(\theta,\varphi) = \sqrt{\frac{(2l+1)}{4\pi} \frac{(l-m)!}{(l+m)!}}
  e^{i m \varphi} P^m_l(\cos(\theta)).
\end{equation}
Some visualizations on the sphere are given in Appendix~\ref{app:ylm}. The angular variables relative to the cluster position $M$ are defined so that $\theta$ is the angle that $(OM)$ makes with the major axis, and $\varphi$ is the angle that $(OM^\prime)$ makes with the intermediate axis, with $M^\prime$ the projection of $M$ in the plane perpendicular to the major axis.
Note that since $\xi_\mathrm{hh}^{\Epsilon}(\ve{r})$ is a real function, ${m < 0}$ coefficients can be  identified with $m > 0$ ones using the relation: $\overline{a_{l,m}} = (-1)^m a_{l,-m}$. Also, $\xi_\mathrm{hh}^{\Epsilon}(\ve{r})$ is invariant under $\pi$-rotations around any of the three axes,  leading to ${a_{l,m} = 0}$ when $l$ or $m$ is odd, and ${a_{l,m} \in \mathbb{R}}$. Accordingly in the following we choose to restrain our harmonic description to $a_{l,m}$ functions for $(l,m) \in \{(0,0), (2,0), (2,2), (4,0), (4,2), (4,4), (6,0)\}$. Higher harmonics quickly become negligible, hence the previous harmonics correspond to a reasonable compromise between restraining the number of coefficients and keeping the most relevant ones to describe $\xi_\mathrm{hh}^{\Epsilon}$. 

Fig.~\ref{ylmDec} shows the results up to 30 Mpc in Lagrangian space (top panels, with $\Delta r = 1.5~\mathrm{Mpc}$) and Eulerian space (bottom panels, with $\Delta r = 0.5~\mathrm{Mpc}$) for three mass bins from left to right. For a given radius, surface integrals on the sphere are computed using a trilinear interpolation of the regular grid of values corresponding to the $\xi_\mathrm{hh}^{\Epsilon}$ function. Note that we make the radial step size of $a_{l,m}$ functions higher for Lagrangian measurements to prevent pixelization effects due to the limited resolution of the simulation.
The monopole $a_{00}$ is directly proportional to the isotropic correlation calculated in Fig.~\ref{2Iso} with, in particular, an exclusion and a bump centered at increasing distances for increasing masses. The chosen normalisation for $Y_0^0$ leads to a factor of proportionality equal to $\sqrt{4\pi}\sim 3.5$.
In Lagrangian space, the monopole is clearly dominant in the exclusion zone. For larger radii, it is exceeded by the quadrupolar component which is a direct evidence for the importance of filamentary clustering (equivalently the lack of clustering in the neighbouring void regions). Higher order contributions also appear with lower amplitude when $l$ grows. This is a signature of the connectivity of the cosmic web with filaments typically bifurcating even more so that the haloes under consideration are massive. This was modelled from first principles and thoroughly investigated in \cite{2018MNRAS.479..973C} where more massive haloes/nodes where shown to be more connected to their environment   {\citep[as was also tentatively measured in numerical simulations -- although with large uncertainties -- in][]{2010MNRAS.408.2163A}}. Typically the low-mass haloes are connected to $\sim 4$ filaments on average while larger mass haloes can reach typically 8-10 connections. This is one reason why more massive haloes excite harmonics of higher orders. 
At low mass, haloes are embedded in one big filament so that two branches emanate on each side and a quadrupole in the correlation function appears. Possibly two more branches are also there but on average they will be less pronounced hence the lower amplitude of the corresponding harmonics. There is no significant contribution beyond. On the other hand, massive haloes are connected to about 8-10 branches of filaments which is why harmonics up to higher order (8 or so) can be excited. Obviously, the low orders dominate (corresponding to the dominant filaments) and higher orders are progressively suppressed as they correspond to fainter ones.

In Eulerian space, gravitational collapse sharpens the filamentary cosmic web. The clustering of haloes is enhanced as they moved along the cosmic web. As can be seen on the bottom panel of Fig.~\ref{ylmDec}, the monopole dominates now over the entire range of separations. This is especially true for the smaller mass haloes, maybe because those in clusters have erased memory of their large-scale environment and therefore tend to reduce the relative impact of higher order harmonics. However, for the more massive bin, the quadrupole is large and reaches more or less the same amplitude as the monopole for all separations. This is probably because larger masses have evolved much less and have kept their initial main connections (although some small filaments may have merged).

\subsection {Oriented stacks with symmetry-breaking}
\label{symSection}

\begin{figure*}
	\centering
		\includegraphics[width=\columnwidth]{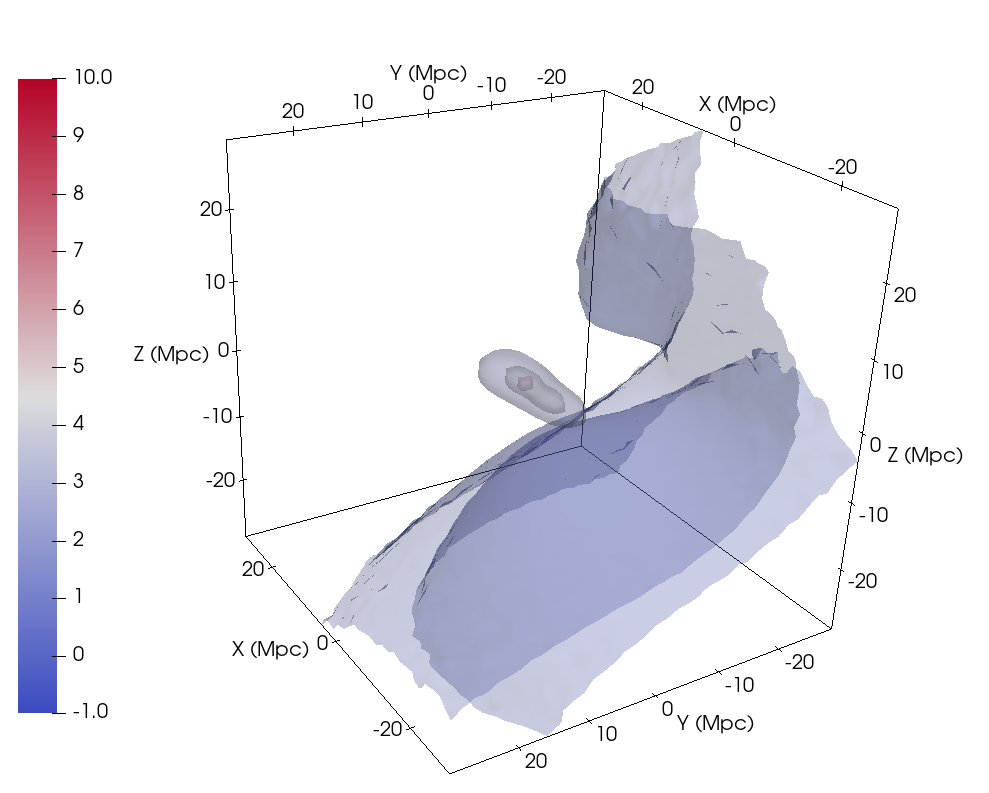}
		\includegraphics[width=\columnwidth]{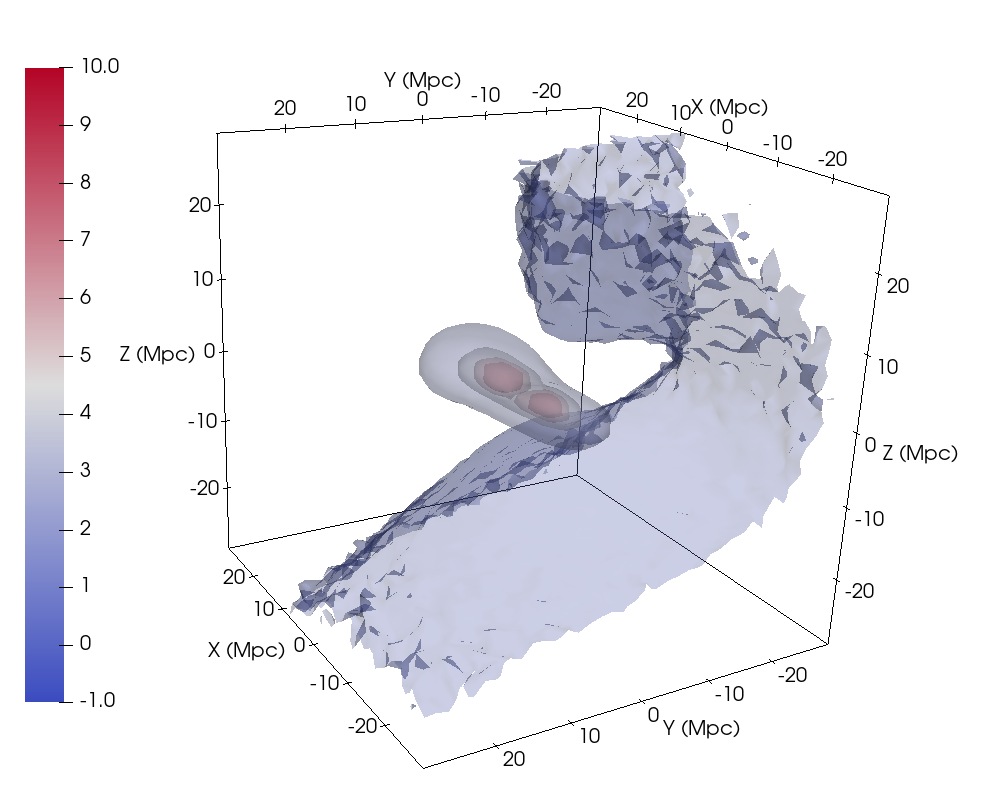}
	\caption{Same as Fig.~\ref{3D-DD} after symmetry breaking in Eulerian space for two mass bins: $1.5\times 10^{12}$--$10^{13}M_\odot$ (left) and $10^{13}$--$10^{14}M_\odot$(right).}
	\label{3D-DD-asym}
\end{figure*}

Previously, we have assumed no preferential orientation (i.e global sign) of the eigenvectors. Those were headless arrows. Hence, one could only consider one eighth of the space to represent the 3D oriented correlation function. This has two advantages: eight times more data per pixel which is useful to reduce numerical noise, and a memory usage divided by 8. However,
there might be some physical reasons to impose an orientation for the basis. For example if one wants to take into account the displacement of haloes between the initial and final positions to compute the correlations, one could impose the eigenvectors of our haloes to be oriented approximately in the direction of the displacement vector which more or less boils down to orientating filaments in the direction of the force (i.e the attractive node). Consider a halo with a displacement vector $\ve{s}$, which in linear theory is proportional to its gravitational acceleration vector. We can add information to the stacking without losing any information by orienting the positive directions of the frame axes to be preferentially pointing towards the gravitational acceleration, in particular by imposing that the three eigenvectors $\{\ve{\hat{e_i}}\}$ of the strain tensor $\Epsilon$ are now directed and not headless: 
\begin{equation}
\forall i \in \{ 1,2,3\}, \ve{s}\cdot\ve{\hat{e_i}} > 0.
\end{equation}
This representation allows us to get some information about how the flow of haloes influences the distribution of haloes around the principal axis of the strain tensors. We refer to this approach as  \textit{dipolar symmetry breaking} in the following.

\begin{figure*}
	\centering
    \includegraphics[width=\textwidth]{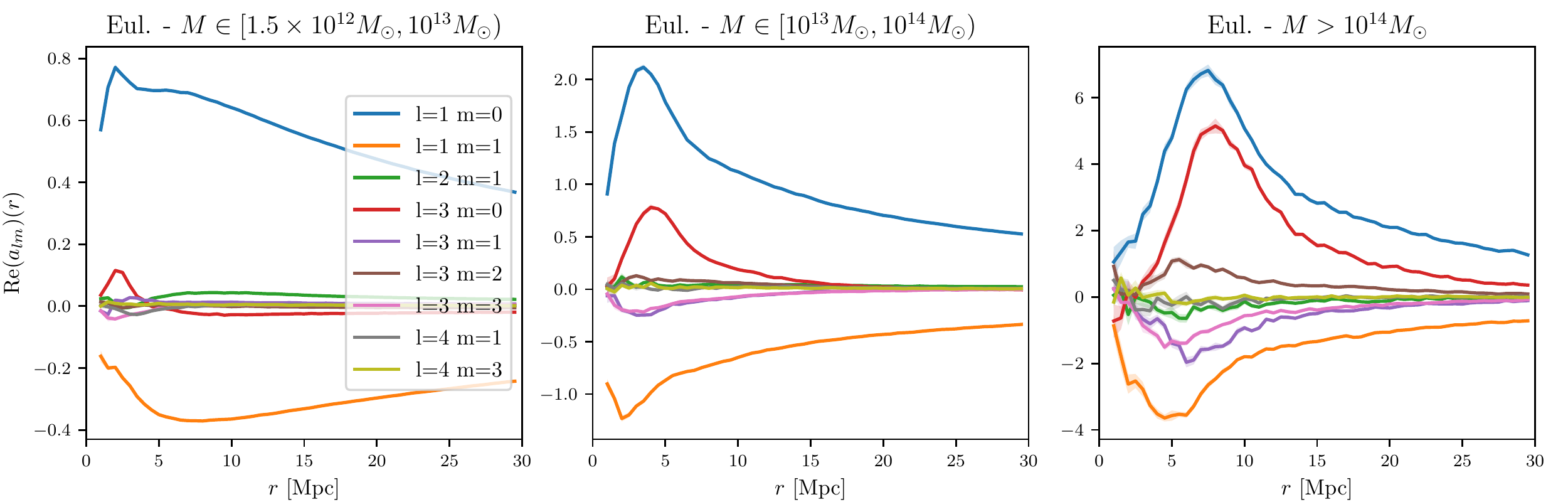}
	\caption{$Y_l^m$ decomposition of $\xi_\mathrm{hh}^{\Epsilon}(\ve{r})$ with symmetry breaking on spheres of various radii.   {We show the real part of the first $a_{l,m}$ coefficients, excluding those corresponding to even $l$ and $m$ values, which are redundant with the ones plotted in Fig.~\ref{ylmDec} by symmetry.} Results in Eulerian space are displayed for different mass bins from left to right as labeled. The legend gives the corresponding $(l,m)$ values.}
	\label{ylmDec-asym}
\end{figure*}

\begin{figure*}
	\centering
    \includegraphics[width=\textwidth]{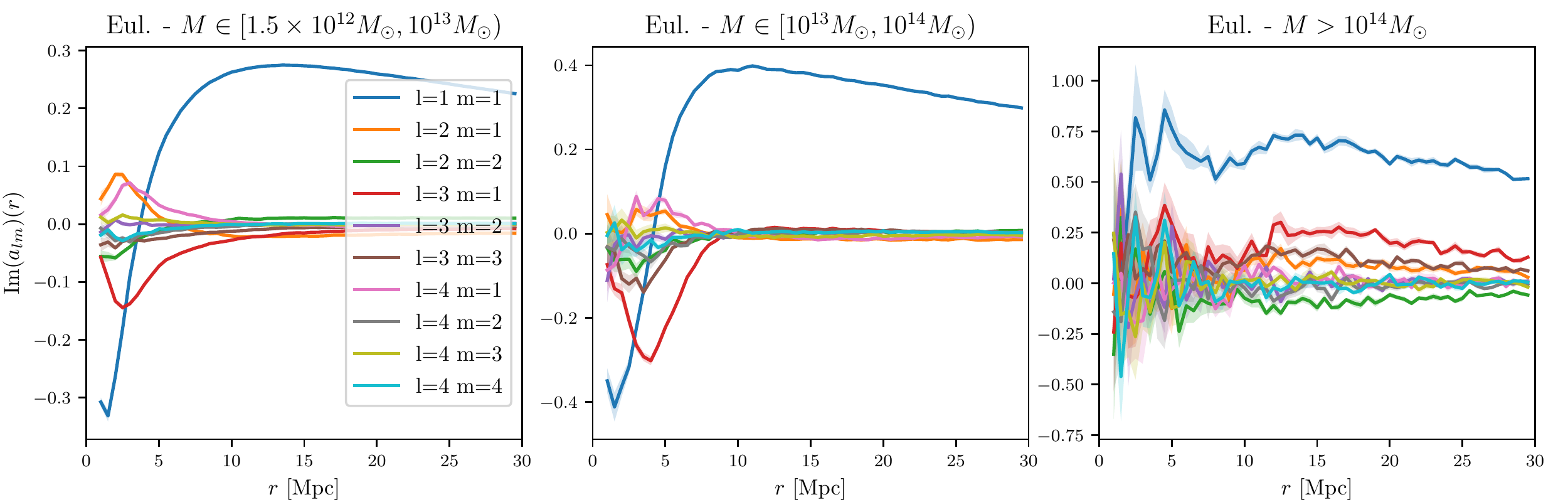}
	\caption{Same as Fig.~\ref{ylmDec-asym} for the imaginary part 
	of the   {$a_{l,m}$} coefficients.   {$m=0$ coefficients have a null imaginary part and are therefore not displayed.}}
	\label{ylmDec-asym-im}
\end{figure*}

Fig.~\ref{3D-DD-asym}
shows the impact of the flow symmetry breaking on the oriented density-density correlation function $\xi_\mathrm{hh}^{\Epsilon}(\ve{r})$. The 3D correlation has been visually blown in the direction of the flow. This asymmetry means that haloes are more likely to cluster in the direction of the flow. On the $a_{l,m}$ coefficients, as seen on  Figs~\ref{ylmDec-asym} and \ref{ylmDec-asym-im}, this therefore excites odd modes that were null before, mainly the ones corresponding to $l=1$ (the real part of even modes are left unchanged and therefore not displayed again here). These $l=1$ modes correspond to the flow being from the voids to the walls, from the walls to the filaments and finally from the filaments to the nodes and simultaneously the clustering being more pronounced in walls, filaments and even more nodes. Hence, the main mode excited is along the filaments therefore it corresponds to the harmonics $(1,0)$, followed by the walls and voids. The amplitude of all modes is larger for larger masses as expected. Interestingly, the relative importance of higher order harmonics, notably $(3,0)$, increases for larger masses.

\subsection{Alignment of the strain tensors}
\begin{figure*}
	\centering
	\includegraphics[width=\textwidth]{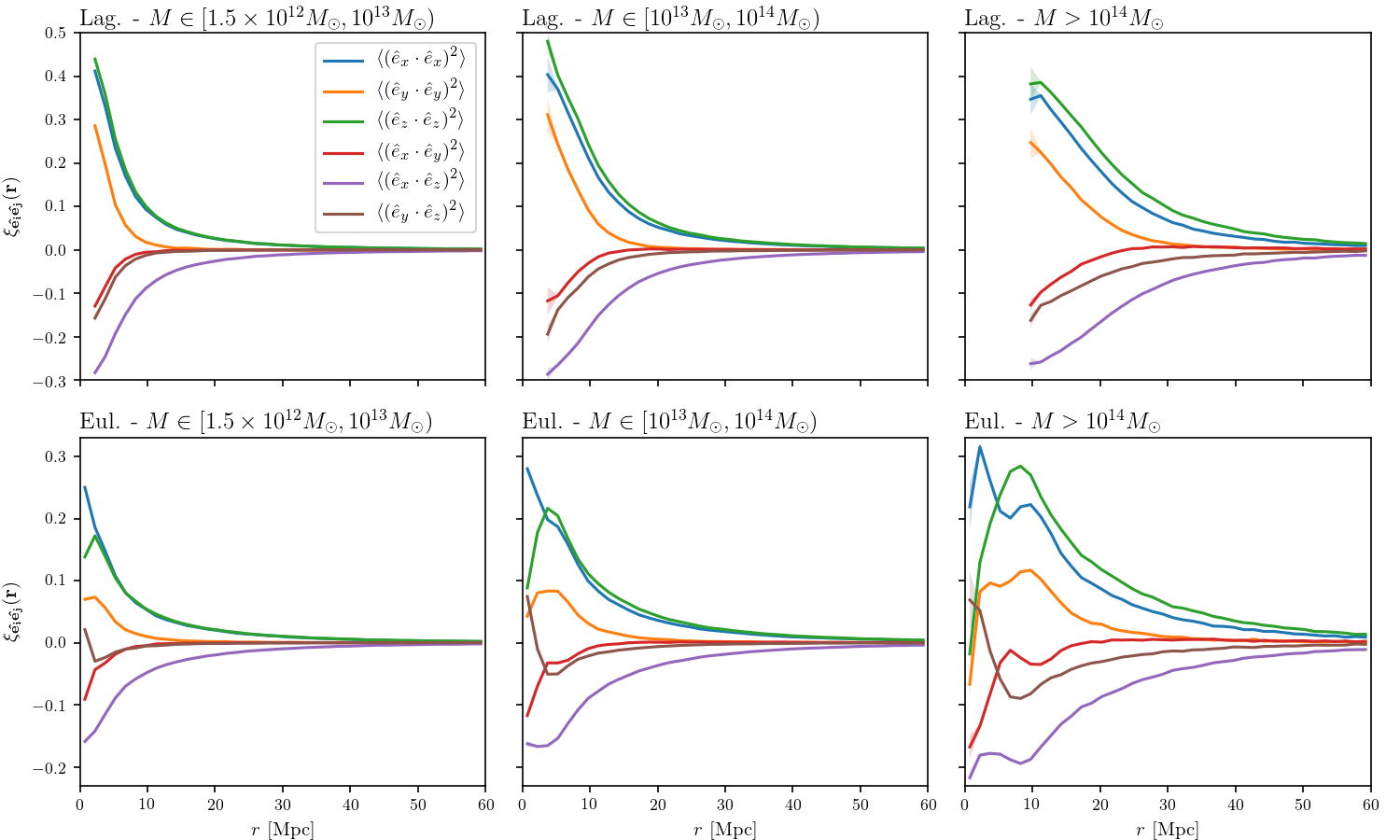}
	\caption{$\xi_{\ve{\hat{e_i}}\ve{\hat{e_j}}}(\ve{r})$ correlation functions quantifying halo intrinsic alignments, as measured in Lagrangian (top panels) and Eulerian (bottom panels) spaces for different mass bins as labelled.}
	\label{axisCorr}
\end{figure*}

Previously we have studied in detail the anisotropic halo clustering by constraining the strain of the target halo, but not using the second halo's strain tensor. We use this information to explore the intrinsic alignment of haloes by describing how the two principal eigenvector bases are correlated as  their separation changes. For this we compute 
 the (isotropic) two-point correlation function defined by 
\begin{equation}
  \xi_{\ve{\hat{e_i}}\ve{\hat{e_j}}}(\ve{r}) = \langle (\ve{\hat{e_i}}(\ve{r}_{h_2}).\ve{\hat{e_j}}(\ve{r}_{h_1}))^2 \rangle-\frac 1 3,
\end{equation}
where $\ve{\hat{e_i}}(\ve{r}_{h_2})$ refers to the normalised eigenvector of the strain tensor of a halo at given position $\ve{r}_{h_2}$ associated with the eigenvalue $\lambda_i (\ve{r}_{h_2})$.

Fig.~\ref{axisCorr} shows the resulting halo alignments in Lagrangian and Eulerian space for various mass bins. If no alignments were detected, one would get $\xi_{\ve{\hat{e_i}}\ve{\hat{e_j}}}(\ve{r}) =0$.
As expected, halo eigendirections do tend to align,  with a stronger alignment for the minor and major axes compared to the intermediate eigendirections. The coherence of these alignments extends to quite large scales, from about 15 to 50 Mpc as halo mass grows, in agreement with the expected coherence of the tidal tensor from linear theory \citep{2009ApJ...705.1469L}   {and recent measurements in N-body simulations \citep{Kurita_2020}}. Interestingly going from Lagrangian to Eulerian space does not erase these alignments although they are slightly suppressed at small scales as anticipated due to nonlinear effects.
As detailed further in Appendix~\ref{sec:peaks}, linear theory predict qualitatively the same signal although the difference seen between minor and major axis, especially at large mass, is a clear signature of the nonlinearities at play, in particular due to the ellipsoidal collapse dynamics.

This result strongly suggests that halo intrinsic alignment may pervade on very large scales and resist the nonlinear evolution.

\subsection{Orientated stacks in projection}
\begin{figure*}
	\centering
	    \includegraphics[width=\textwidth]{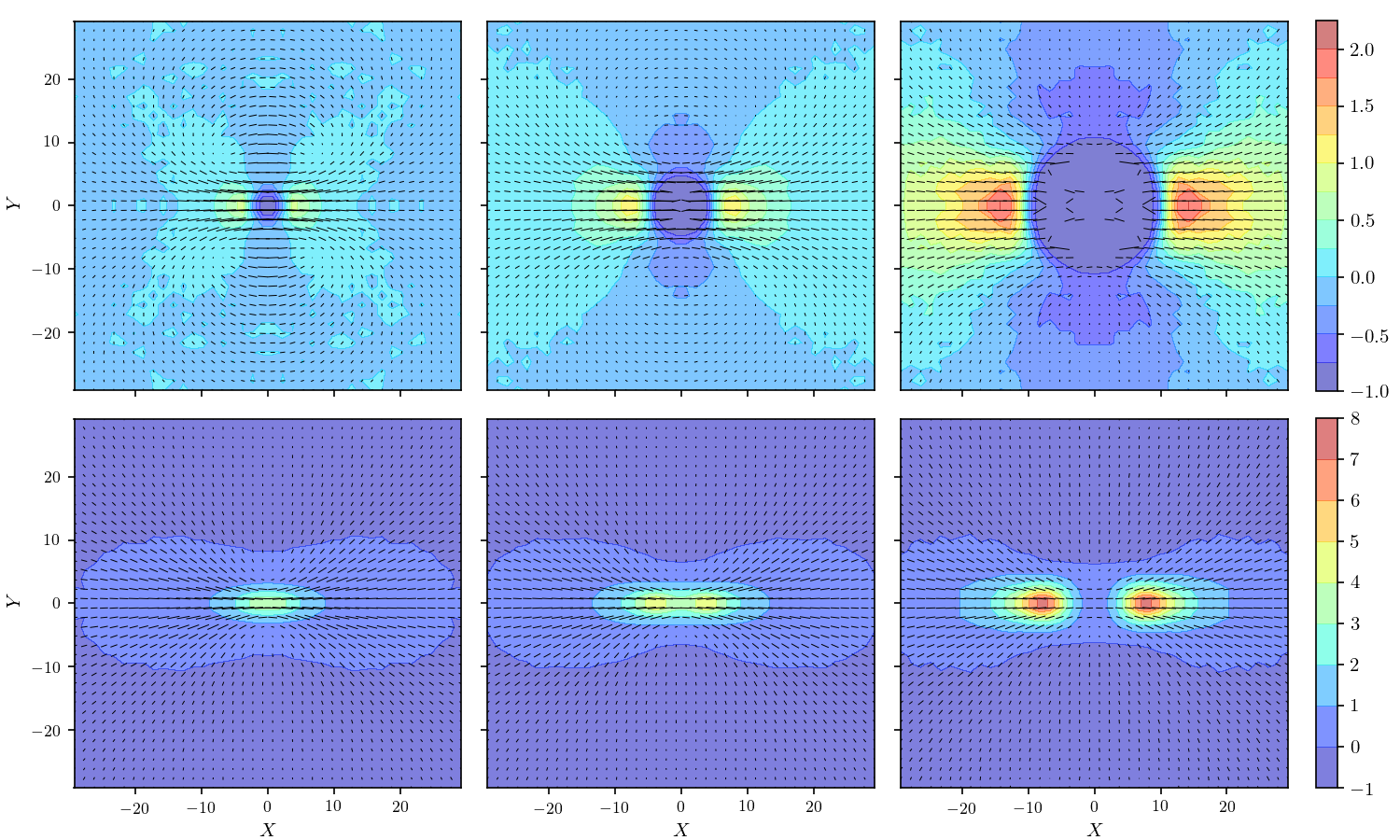}
	\caption{Oriented correlations from 5 Mpc thick projections of the peak patch simulation: the colour map represents the oriented halo density correlation function  $\xi_\mathrm{hh}^{\Epsilon /\mathrm{2D}}(\ve{r})$ while the bars show the average major eigendirection of surrounding haloes. The $X$ axis is the direction of the major axis of the central halo, and the $Y$ axis follows its minor axis. Top panels are in Lagrangian space while bottom panels are in Eulerian space. Three mass bins are considered from left to right: $M \in {[ 1.5 \times 10^{12} M_{\odot}, 10^{13} M_{\odot} ]}$; $M \in {[ 10^{13} M_{\odot}, 10^{14} M_{\odot}]}$; $M \in {[ 10^{14} M_{\odot}, +\infty [}$.  }
	\label{2Dproj}
\end{figure*}

Visualizations of 3D correlations in the static format of the 2D page do not convey as much information as one would like, making a case for 2D projections. This is also relevant for observations, which can deliver 2D maps, possibly with redshift foliation of the maps (photometric clustering, cosmic shear, intensity mapping, CMB, etc). Hence we now show projections of 3D results into 2D.
For that purpose, we choose an arbitrary axis of the simulation, let us call it $(Oz)$, and define some $z$-slices of the simulation that are then projected onto the same plane. 
With two-point correlation functions,
for a given $z$-slice, the projection of halo positions is simply achieved by ignoring the $z$-coordinates and using directly the $(x,y)$ coordinates as the projected coordinates of the objects. To calculate a strain-tensor-oriented correlation, we also need to compute the 2D projections of the strain tensors on the $(xOy)$ plane,  which is simply
\begin{equation}
  \Epsilon_{\rm 2D} =
  \begin{pmatrix}
   \epsilon_{xx} & \epsilon_{xy}\\
   \epsilon_{yx} & \epsilon_{yy}
  \end{pmatrix}.
\end{equation}

Using these projections within 5 Mpc slices, we then compute the oriented density-density correlation function in 2D $\xi_\mathrm{hh}^{\Epsilon /\mathrm{2D}}(\ve{r})$
using the same techniques as for the 3D function. Hence, we estimate the distribution of neighbouring haloes in the vicinity of a central halo with fixed orientation. In addition, we also measure in the same frame the average major eigendirection of the neighbours (set by their strain tensor) in order to get a clear visualisation of the shape alignments in the surroundings of a central halo. Fig.~\ref{2Dproj} shows the results of these calculations for several mass bins. The background coloured density map represents  $\xi_{\rm hh}^{\Epsilon /\mathrm{2D}}$, while headless arrows show the average direction of the major axis of surrounding haloes, and their length is proportional to the difference between the eigenvalues.
Here, everything is computed within the frame of the central halo, the $X$ axis being its major direction, and the $Y$ axis its minor direction. There is no symmetry breaking which is why the plots are reflection symmetric with respect to the two axes.

These projections show clearly the exclusion zone in dark blue growing from small to larger masses with a clear anisotropy as it is squashed along the minor axis of the central halo. The clustering of neighbours peaks again (in yellow/red) at an increasing separation for larger masses. It does not happen  in all directions but specifically along the major axis of the central halo. This anisotropy is even more pronounced in Eulerian space (bottom panels). All those results are nothing but
 a 2D visualisation of the same phenomenon already observed for 3D correlations (see Fig.~\ref{2Iso}) with a clear filament appearing along the major direction and two voids along the minor axis.
 
 Looking at the black bars which represent the stacked orientation of neighbouring haloes, a clear alignment of their shape is detected, in particular close to the X axis. This shows an average alignment of haloes along the major direction (halo shapes tend to point towards each other) and for greater scales ($\geq30$~Mpc, beyond the boundary of the plot) than what we already observed in isotropic settings like the one shown in Fig.~\ref{axisCorr}. 
 Tidal alignments seem therefore  to occur on very large scales, even larger than the mean correlation length of the tidal field. This is a consequence of the so-called biased clustering effect \citep{1989MNRAS.237.1127C}: haloes do not form anywhere in the Universe  but along filaments which are very special places where the field is coherent. This induces very strong and long-range halo correlations along the cosmic web.

\begin{figure*}
	\centering
			\includegraphics[width=\textwidth]{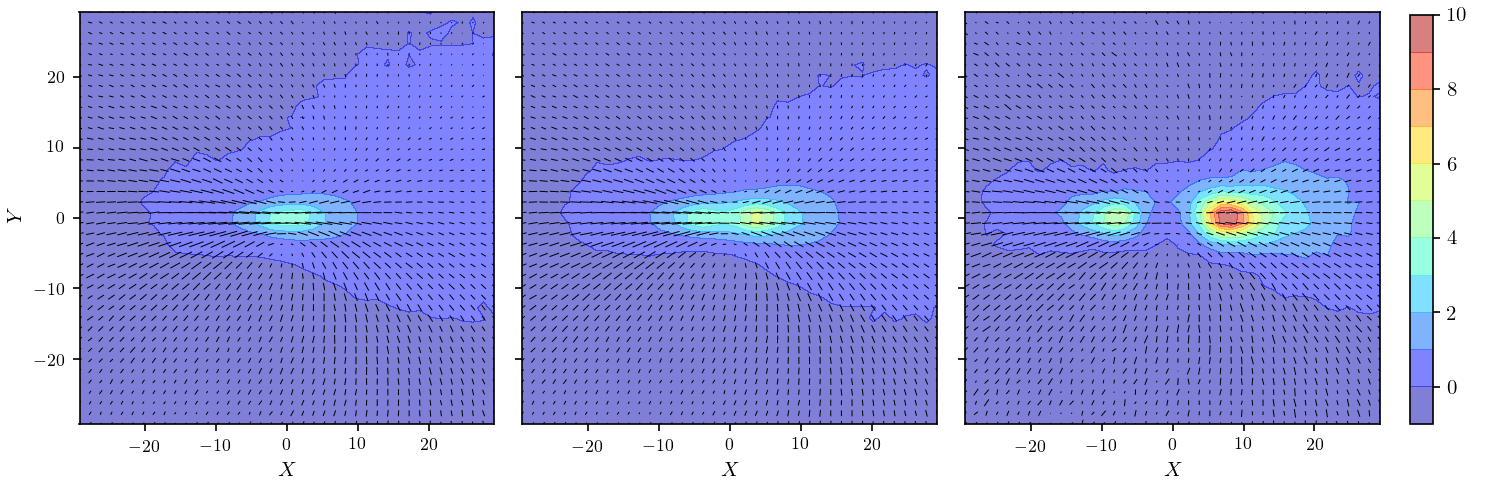}
	\caption{Same as the bottom panels of Fig.~\ref{2Dproj} with symmetry breaking in Eulerian space.}
	\label{2Dproj-SB}
\end{figure*}

Finally, we also compute the same projected correlations after symmetry breaking (taking into account the orientation of the flow).
The result is displayed in Fig.~\ref{2Dproj-SB}. As expected, the clustering becomes dissymetric with many more neighbours in the direction of the flow. The orientation of the neighbours is however relatively unaffected by the symmetry breaking procedure.

\section{Discussion and conclusion}
\label{sec:conclusion}

In this work, the peak patch picture was used as a model for halo formation and evolution.
Based on this model, the anisotropic halo clustering was first investigated. The anisotropy could be revealed thanks to an oriented stack in the frame of the halo's strain tensor whose orientation is very correlated with the surrounding cosmic web. Haloes were shown to cluster mostly in the direction of the filaments as expected. This effect was further quantified using an harmonic decomposition of the three dimensional correlation function, which allowed us to make the connection with the so-called cosmic web connectivity modelled in \cite{2018MNRAS.479..973C}. In addition, we also implemented a symmetry breaking procedure where the direction of the local flow is accounted for. As expected haloes cluster more in the direction of the flow.

Furthermore, the correlations between halo shapes as traced by their initial strain were quantified both in 3D and in projection. It was shown that those correlations exist on very large scales and are highly anisotropic. Along filaments, they extend to scales as large as 50 Mpc and more. 

In order to provide theoretical insight onto those results, first principle calculations based on the hierarchical peak theory of Gaussian random fields can be performed. Such predictions for the oriented clustering of peaks in Gaussian random fields are provided in Appendix~\ref{sec:peaks} and are shown to  lead to the same qualitative picture as for the peak patch picture described in this paper. However, one should bear in mind that the peak theory description, although elegant since it can be fully written and derived from first principles, cannot model accurately the details of halo formation and alignments which require a more complex modelling of their collapse, displacements and nonlinear exclusion effects which are provided by the mass-peak patch model used here.

Of course it is quite likely that other elements of nonlinear gravitational evolution and the complex baryonic processes (such as feedback from active galactic nuclei, galactic winds, supernovae explosion, turbulence, etc) associated with structure formation may decrease the initial alignments modelled in this article. However the work presented here allows us to understand more clearly one of the building blocks of halo intrinsic alignments: the large-scale tidal coherence. On smaller scale, this ingredient should be convolved with the complex response function of galaxies to this underlying tidal correlations. Hence, it would be of great interest to develop a hybrid framework combining the large-scale tidal alignments modelled in this paper with galaxy response functions that could be measured in hydrodynamical simulations. This is however beyond the scope of this paper and will be studied elsewhere.

  {Let us also note that in this study we have not considered the build up of halo spin which also tends to align with the cosmic web \citep{codisetal12,2013ApJ...762...72T,2017MNRAS.468L.123W,2018MNRAS.473.1562W,2018MNRAS.481..414G}. This would require a better understanding of the shape of protohaloes, although a simple proxy based on initial peak shape following \cite{ATTT} could easily be used. This will be the subject of future works.}

\section*{Acknowledgements}
  {We thank the anonymous referee for a helpful report.}
These calculations were performed on the Niagara supercomputer \citep{niagara} at the SciNet HPC Consortium. SciNet is funded by: the Canada Foundation for Innovation under the auspices of Compute Canada; the Government of Ontario; Ontario Research Fund - Research Excellence; and the University of Toronto.
Part of the work has also made use of the Horizon Cluster hosted by Institut d'Astrophysique de Paris.
We thank Stephane Rouberol for running smoothly this cluster for us.
This work is partially supported by Fondation MERAC, the the Natural Sciences and Engineering Research Council of Canada (NSERC) and the Canadian Institute for Advanced Research.
BRSB thanks CITA for supporting his internship in 2017 when this work was initiated.
We also thank Fran\c{c}ois Boulanger and Fran\c{c}ois Levrier for their support and are grateful to Marcelo Alvarez for fruitful discussions and inputs. JRB also thanks Zhiqi Huang, Andrei Frolov, Connor Bevington, Martine Lokken, and Ronan Kerr for developing applications of the methodology described here to a variety of projected 2D observables derived from large datasets, e.g., to cosmic microwave primary anisotropies in the Planck and ACT maps and the  primordial curvature oriented stacks derived from them, to superclustering in the thermal Sunyaev Zeldovich effect, to Line Intensity Mapping, and to Galactic dust filamentariness in the Planck maps. 

\section*{Data availability}
The data underlying this article are available in the article.

\bibliographystyle{mnras}
\bibliography{bibli}

\appendix

\section{Oriented stacks of peaks in Gaussian random fields}
\label{sec:peaks}

\begin{figure*}
	\centering
	\includegraphics[width=0.95\columnwidth]{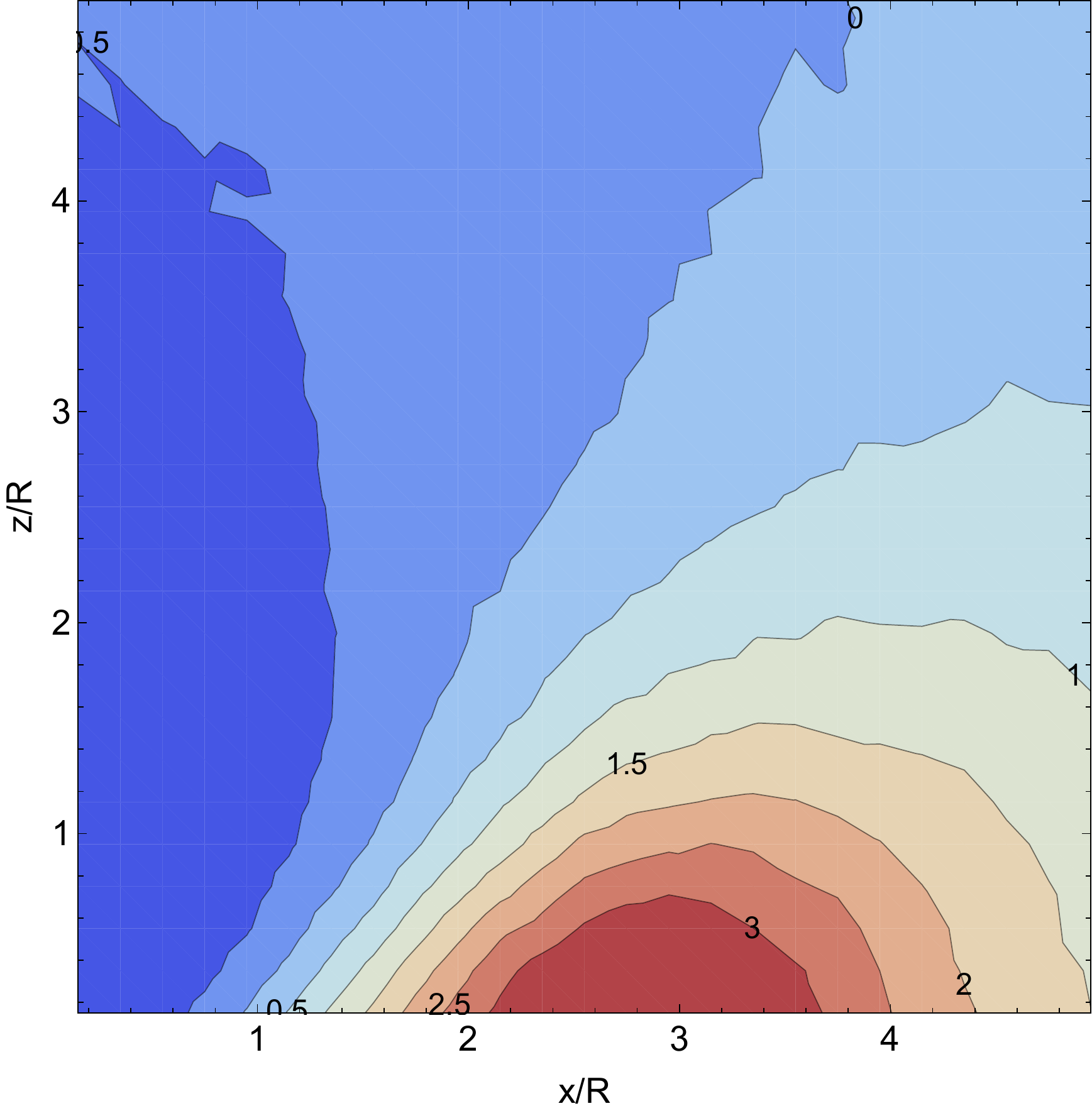}
	 \includegraphics[width=1.01\columnwidth]{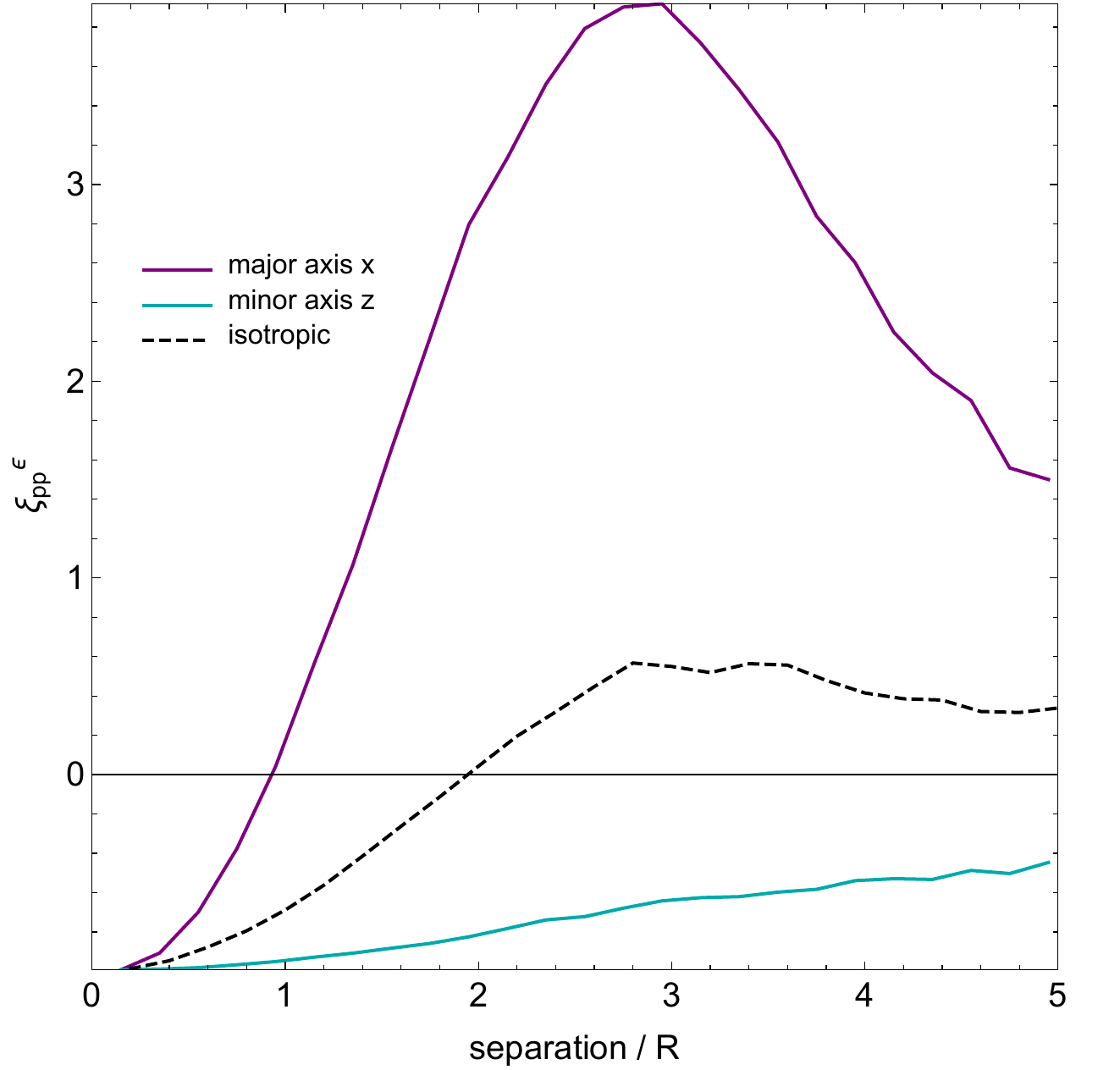}
	\caption{Oriented correlation function of Gaussian peaks with height between 1.5 and 2.5 sigma for a power-law power spectrum with spectral index $n_s=-2$. Left-hand panel: Correlation function in the plane $y=0$. Right-hand panel: Correlation function along the minor ($z$-axis) in cyan and major eigendirection ($x$-axis) of the shear in purple as labelled. For comparison, the isotropic correlation function is shown with dashed lines.}
	\label{fig:pk-pk-nu2}
\end{figure*}

Let us consider a statistically homogeneous and isotropic Gaussian random field $\delta_{\rm L} (\ve{r})$ (the linear density contrast) with zero mean and 
power spectrum $P_k$, where $k$ is the magnitude of the comoving wavenumber.
In this section, we will investigate how peaks of this matter density field cluster in the frame of the strain tensor. This is meant to be a toy model to explain the oriented stacks of haloes measured in peak patch simulations and described in the main text.

\subsection{From mass-peak patches to overdensity peaks}

As discussed and shown in BM1, in spite of the complexity of exclusion a surprisingly simple semi-analytic approximation for the number density of hierarchical mass-peaks works reasonably well compared with the full hierarchically merged numerical results, and we will use that for the illustration of how the main effects of exclusion and large scale anisotropic clustering emerge. The homogeneous ellipsoid dynamical model was described in Section~\ref{sec:PP}. The collapse along all 3-axes to a final equilibrium can be translated to initial condition (linear) space, through a critical relation between $\delta_{\rm L}$ and the anisotropic eigenvalues, $\nu e_v$ and $\nu p_v$: $\delta_{\rm L} = \delta_{\rm L, crit} (e_v, p_v)$.

At a given mass with top hat radius $R_c$, $M_c \propto R_c^3$, at the redshift in question, there are probability distributions of  $\delta_{\rm L}(\ve{r}_c\vert R_c)$, $e_v(\ve{r}_c\vert R_c)$ and $p_v(\ve{r}_c\vert R_c)$ with mean values $\langle \delta_{\rm L} \rangle (\ve{r}_c\vert R)$, $\langle e_v \rangle (\ve{r}_c\vert R)$ and $\langle p_v \rangle (\ve{r}_c\vert R)$ and dispersions about those means, 
The BM1 semi-analytic approximation neglects $p_v$, replaces $e_v$ by $\langle e_v \rangle (\ve{r}_c\vert R_c)$ and uses $\delta_{\rm L} (\ve{r}_c\vert R_c)= \delta_{\rm L, crit}(\langle e_v\rangle)$. This also results in a function giving the cluster top-hat scale $R_c (\delta_{\rm L, crit}, \langle e_v \rangle)$. As the mass drops $\langle e_v \rangle $ and $\delta_{\rm L, crit}$ rise, the mass-peaks have greater initial anisotropy than high mass peaks. With this approximation mass is on average related to $\delta_{\rm L, crit}$, so the  mass-peak finding criterion, $\nabla M =0$, translates into a density-peak finding criterion, $\nabla \delta_{\rm L} =0$. The position and mass delta function relations, $\ve{r}=\ve{r}_c$ and  $M=M_c$,  can be transformed into delta function relations for the $\delta_{\rm L}$:

\begin{eqnarray}
    \delta(\ve{r} -\ve{r}_c) \delta (\ln M-\ln  M_c) \nonumber\\
 &&\hskip -4cm  =
    {\rm det}\vert \nabla \nabla \ln M \vert \delta(\nabla \ln M) \delta(\ln M -\ln M_c)\nonumber\\
  &&  \hskip -4cm \rightarrow {\rm det}\vert \nabla \nabla \delta_{\rm L} \vert \delta(\nabla \delta_{\rm L}) \vert \frac{{\rm d}\delta_{\rm L}}{{\rm d}\ln M} \vert  \delta (\delta_{\rm L}  -\delta_{\rm L, crit}) \\
  &&\hskip -4cm 
  = n_{{\rm p,op}} (\nu \vert R_c) \vert \frac{{\rm d}\nu } {{\rm d}\ln M} \vert  \delta (\nu -\nu_{\rm  crit})\, .  \nonumber
\end{eqnarray}
Here $n_{\rm p,op}(\nu) {\rm d}\nu  $ is the conventional BBKS density-peak operator as a function of peak height $\nu = \delta_{\rm L}/\sigma_0$. 
(Whether one uses the conventional $ \rho / \bar{\rho}-1$ or $\ln \rho / \bar{\rho}$ for $\delta_{\rm L}  $ does not matter here in this fully linear regime.)

The mass-Peak Patch algorithm, and halo finding generally, is based on top hat filtering. The original BBKS peak description was more focused on Gaussian filtering, hence to relate to the well known BBKS paper we shall adopt Gaussian filtering in this appendix. Typically $R_{\rm TH} =R_c \approx 2R_{\rm G}$ is found in full peak patch simulations, with a small variance around it, and this also follows from the analytic theory. Instead of determining $R_{\rm TH}$, BBKS adopted a relation $M=\bar{\rho}_0 (2\pi)^{3/2} R_{\rm G}^3$ that used the Gaussian volume which gives  $R_{\rm TH}=1.56R_{\rm G}$ rather than the peak patch $\approx 2R_{\rm G}$. For Gaussian filtering ${\rm d}\delta_{\rm L} /{\rm d}\ln M= R_{\rm G}^2 \nabla^2 \delta_{\rm L} /3 = R_{\rm G}^2 \sigma_2 I_1 /3$, where the spectral parameter $\sigma_2^2=\left\langle(\nabla^2 \delta_{\rm L})^2\right\rangle$ and the dimensionless curvature variable $I_1=\nabla^2 \delta_{\rm L}/\sigma_2$
are familiar from BBKS. 
The approximate semi-analytic relation for the hierarchical peaks operator $n_{\rm hpk,op} (\ve{r},\ln M_c ){\rm d}^3\ve{r}{\rm d}\ln M $ is then expressible in terms of the BBKS peak density operator $n_{\rm p,op}(\ve{r}\vert R_G,\nu){\rm d}^3 \ve{r}{\rm d}\nu $, where $\nu=\delta_{\rm L}/\sigma_0$ and  $\sigma_0^2=\langle \delta_{\rm L}^2\rangle$, by
 \begin{equation}
n_{\rm hpk,op} (\ve{r}_c,\ln M_c , q_c) = \frac{\sigma_2R_{\rm G}^2}{3\sigma_0} I_1   n_{\rm p,op}(\ve{r}_c,\nu_{{\rm crit},c},q_c),
\end{equation}
where $q_c$ can be an additional internal property of the haloes.

In BBKS the determination of the one-point distribution $\langle n_{\rm pk}\rangle(\nu) d\nu$ of overdensity peaks was emphasized, but estimate of object numbers was done by integrating over a ${\rm Prob}(\nu_{th} \vert \nu)$ selection function of peaks above a threshold $\nu_{th}$.  Now with the peak patch theory we have a selection function ${\rm Prob}(\nu_{\rm th} \vert \nu)= \delta (\nu -\nu_{\rm crit}(M_c))$. Selection functions may be a simple way to extend the analytics described here to take into account the spread of $\nu$ about $\nu_{\rm crit}(M_c)=\langle \delta_{\rm L} \vert M_c\rangle/\sigma_0(R_c)$ for given $M_c$. However all the anisotropic variables would have their own selections, and all would have to be calibrated by the full peak patch simulations, especially relevant if clustering functions are the target.  In the next section we allow a range of $\nu$ for a specified mass which can be thought of as motivated by the spread about the mean $\nu_{\rm crit}(M_c))$. 

In these density distribution functions we carry along variables $q$, which include the original ones defined below, plus others fundamental to anisotropy and alignment studies, in particular $\epsilon_{{\rm L},ij}^\prime$ included in BM1 and BKP but which were not included in the original BBKS treatment. Just as $\nu$ and $I_1$ are correlated ($\langle \nu I_1 \rangle =-\gamma, \ \gamma = \sigma_1^2/\sigma_0\sigma_2$), so all components of anisotropic strain $\epsilon^\prime_{{\rm L}, ij}$ and anisotropic curvature $\nabla^2 \epsilon^\prime_{{\rm L},ij} =[\nabla_i \nabla_j \Tr(\epsilon_{\rm L})]^\prime$ are correlated. In particular  $\langle I_1 \vert \nu \rangle \approx- \gamma \nu$ for large $\nu$.

One might wonder how elements of exclusion enter into this semi-analytic framework since no exclusion operator dependent on higher mass peaks enter into this formula. Since peaks of the linear density field must be exactly $\nu_{\rm crit}\sigma_0 (R_c)$, it is not that likely that there will be other peaks within radius $R_c$ satisfying exactly this criterion. There are semi-analytic approximations to half and full exclusion which are integrals over higher masses of $n_{\rm pk} \langle I_1 \vert \nu \rangle $ that give an idea of what the further exclusion acting on the mass function is like, given the restrictions associated with the first upcrossing peaks through $\nu_{crit}$. 

Our target here is to explore semi-analytically the $n_{\rm hpk}\propto n_{\rm p}$ approximation described above to make the two-point correlation of hierarchical peaks, 
\begin{equation} \langle n_{\rm hpk,op}(\ve{r}_1, R_1, q_1\vert {\cal C}_1)n_{\rm hpk,op}(\ve{r}_2, R_2,q_2\vert {\cal C}_2)) \rangle , 
\end{equation}
tractable analytically.   

\subsection{Gaussian statistics of the field and its derivatives}

We now expand the BBKS treatment of $n_{\rm p,op}(\ve{r}_c,\nu_{c},q_c){\rm d}^3\ve{r}_c{\rm d}\nu_c\rm{d}q_c$ to include the $q_c$. Let us focus on the field $\delta_L$ defined above smoothed over a scale $R$ by a filter function $W$ 
 \begin{equation}
  \delta_{\rm L}({\bf r},R) = \int \frac{{\textrm d}^3{\bf k}}{(2\pi)^3} \delta_L({\bf k}) W(kR)
  \mathrm{e}^{i{\bf k}\cdot{\bf r}}\,.
  \label{eq:delta-definition}
\end{equation}
In peak patches, the top hat form of $W$ is used. BBKS emphasized more the Gaussian form, which balances the spread in $\ve{k}$ space of the Fourier transform of $W$ and the spread of $W$ in $\ve{x}$ space. 

Since we have translated the $M=M_c$ criterion into a $\delta_{\rm L}$ criterion using mean-field results, the mass-peak statistics are determined by the distribution of $\delta_{\rm L}$ and its first and second derivatives (to define peaks using a constraint of zero gradient and negative curvatures). In addition, we will need the anisotropic parts of the strain tensor, or equivalently the tide $\nabla_i \nabla_j \Phi_N$, the second derivative of the gravitational potential $\Phi_N$, since in linear theory it is proportional to the strain, as noted in Section~\ref{sec:PP}. 

We shall mostly follow the normalized variable convention used in  
\cite{pogosyan/pichon/etal:2009,ATTT}:  $\phi_{ij} = -\epsilon_{{\rm L}, ij} /\sigma_0$, $x=\Tr(\phi_{ij})={\delta_{\rm L}}/{\sigma_0 }$, $x_{k}\equiv {\nabla_k \delta_{\rm L}}/{ \sigma_1}$,
and  $x_{kl}\equiv {\nabla_k \nabla_l \delta_{\rm L}}/{\sigma_2}$. The normalizations use their respective  variances 
\begin{equation}
\sigma_i^2(R)\equiv \frac{1}{2\pi^2}\int_0^\infty {\textrm d}{k} k^2 P_k(k) k^{2i} W^2(k R)\,.
\label{eq:defsigi}
\end{equation}
Note that BBKS used $\zeta_{ij} = - x_{kl}$ and $\eta_i=x_i$ and $\nu$ for $x$.  

The catalogue of object information we are carrying is therefore ${\cal C} =\{
x, \phi_{ij}, x_{ij}, x_i\}$, which amounts to $15= 6+6+3$ variables in all. The displacement $\nabla^{-2} \nabla_i \Tr(\epsilon_{\rm L})$ adds 3 extra variables to the list if we are interested in the linear velocity distribution - the velocities of peaks are reduced because of the correlation of the displacement and $x_i$.  In BM1 the general numerical peak patch catalogue  also carried along the binding energy, which is a linear quantity. So the standard peak patch catalogue-object list has 20 variables, plus the position $\ve{r}_c$ and mass scale $M_c$. 
 
The joint two-point probability distribution function (PDF hereafter) of the set of 15 normalized fields  $\boldsymbol{X}=\{\phi_{ij},x_{ij},x_i\}$ and $\boldsymbol{X'}=\{\phi'_{ij},x'_{ij},x'_i\}$ 
at two prescribed comoving locations separated by a distance 
$\ve{r}$ is ${\cal P}(\boldsymbol{X},\boldsymbol{X'})$. For Gaussian initial conditions, this joint PDF is the multivariate Normal
\begin{equation}
{\cal N}(\boldsymbol{X},\boldsymbol{X'})= \frac{\exp\left[-\frac{1}{2}
\left(\begin{array}{c}
 \boldsymbol{X} 
\\
\boldsymbol{X'} 
 \\
\end{array} \right)^{\rm T}
 \cdot
  \mathbf{C}^{-1}\cdot 
  \left(\begin{array}{c}
 \boldsymbol{X} 
\\
\boldsymbol{X'} 
 \\
\end{array} \right) \right]}{{\rm det}|\mathbf{C}|^{1/2} \left(2\pi\right)^{\rm 15 }} \, , 
\label{eq:defPDF}
\end{equation} 
where $\mathbf{C}_{0}\equiv \langle  \boldsymbol{X}\cdot \boldsymbol{X}^{\rm T} \rangle$ 
and  
$\mathbf{C}_{\times}\equiv \langle  \boldsymbol{X}\cdot \boldsymbol{X'}^{\rm T} \rangle$ are the diagonal
and off-diagonal components of the covariance matrix
\begin{equation}
\quad \mathbf{C}=\left(
\begin{array}{cc}
\mathbf{C}_{0} &\mathbf{C}_{\times}
\\
\mathbf{C}_{\times}^{\rm T}  &\mathbf{C}_{0} 
 \\
\end{array}
\right)\,.
\end{equation}
Because of statistical homogeneity the off-diagonal covariance depends only on the separation vector $\boldsymbol{r}$. 
Equation~(\ref{eq:defPDF}) is the only ingredient needed to compute the expectation value of any quantity involving the strain tensor, the density gradient and second derivatives.

\subsection{Peak clustering}

\begin{figure}
	\includegraphics[width=\columnwidth]{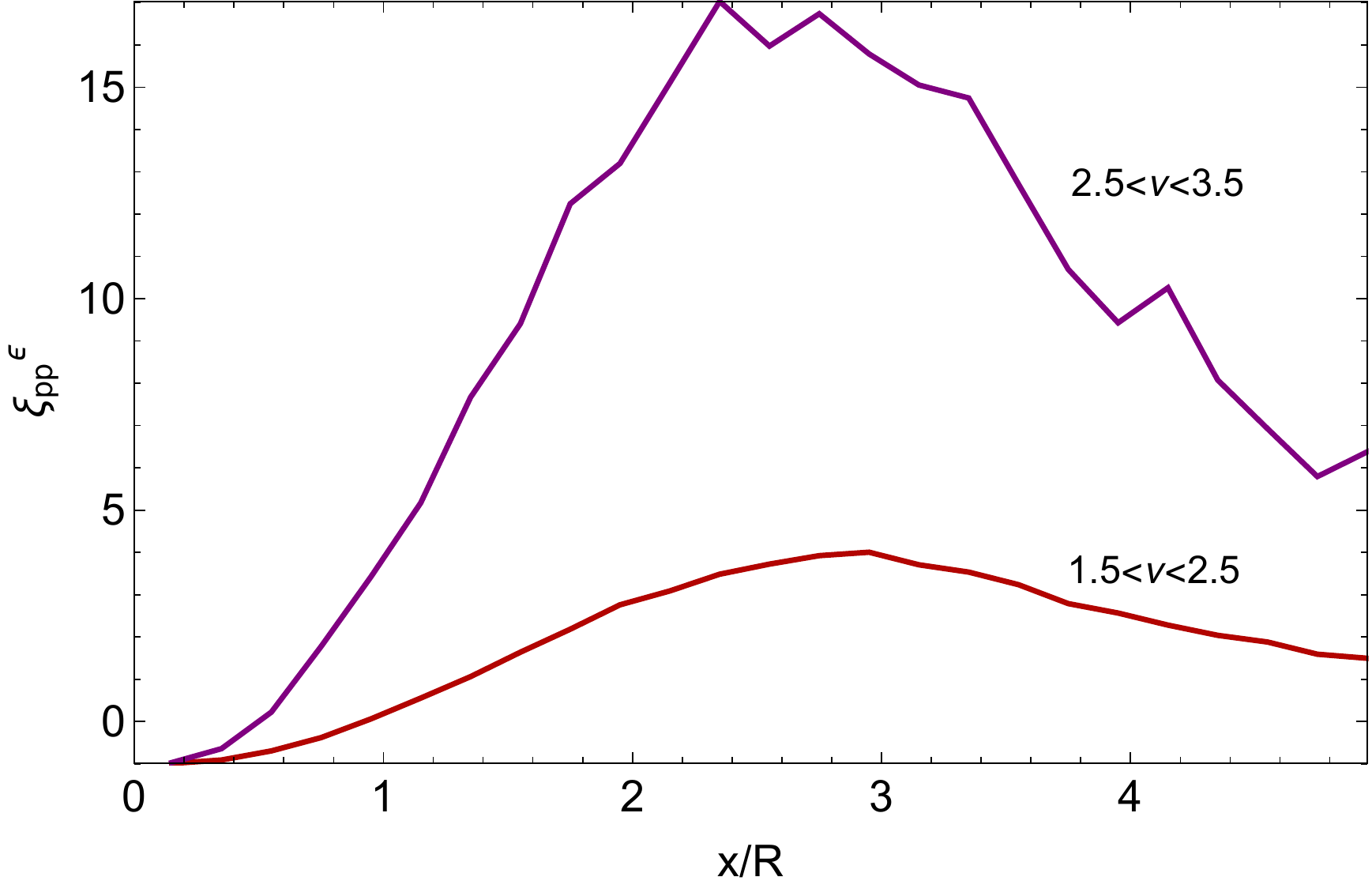}
	\caption{Correlation function of Gaussian peaks of height between 1.5 and 2.5 sigma in red and 2.5 and 3.5 in purple for a power law power spectrum with spectral index $n_s=-2$. }
	\label{fig:pk-pk-nu}
\end{figure}

In particular, equation~(\ref{eq:defPDF}) can be used to compute the two-point correlation $\xi_{\rm pp}(r,\nu)$ of overdensity peaks as a function of $\nu$ separated by $r$ which is given by
\begin{equation}
\label{eq:xicrit}
1+\xi_{\rm pp}(r,\nu)=
 \frac{
\big\langle n_{\rm p}(\boldsymbol{X})  n_{\rm p}(\boldsymbol{X'}) \big\rangle}
{\big\langle n_{\rm p}(\boldsymbol{X}) \big\rangle^2}   \,,
\end{equation}
where the Klimontovich or ``localized'' density of peaks in three dimensions reads \citep{Kac1943,Rice1945,bardeen/bond/etal:1986}
\begin{equation}
\label{eq:ncrit}
n_{\rm p}(\boldsymbol{X})\!=\!-\!\left(\!\frac{\sigma_2}{\sigma_1}\!\right)^\text{3}\!
{\rm det}(x_{ij})\delta_{\rm D}(x_i)\delta_{\rm D}(x-\nu)\Theta_{\rm H}(-\lambda_3) \,,
\end{equation}
with $\lambda_1<\lambda_2<\lambda_3$ the eigenvalues of the density hessian matrix, $\delta_D$ the Dirac delta function and $\Theta_H$ the Heaviside step function.
 The 
dimensional factor $(\sigma_1/\sigma_2)^3$ ensures that the ensemble average 
\begin{equation}
\big\langle n_{\rm p}(\boldsymbol{X})  \big\rangle
\!=\! \int\!{\rm d} \boldsymbol{X} \,
 n_{\rm p}(\boldsymbol{X})   {\cal P}(\boldsymbol{X})
,
\end{equation}
appearing in the denominator of equation~(\ref{eq:xicrit}), is the mean number density of peaks in the band $\nu$ to $\nu+{\textrm d}\nu$ for a given mass (Gaussian filtering scale).

\begin{figure*}
	\includegraphics[width=\columnwidth]{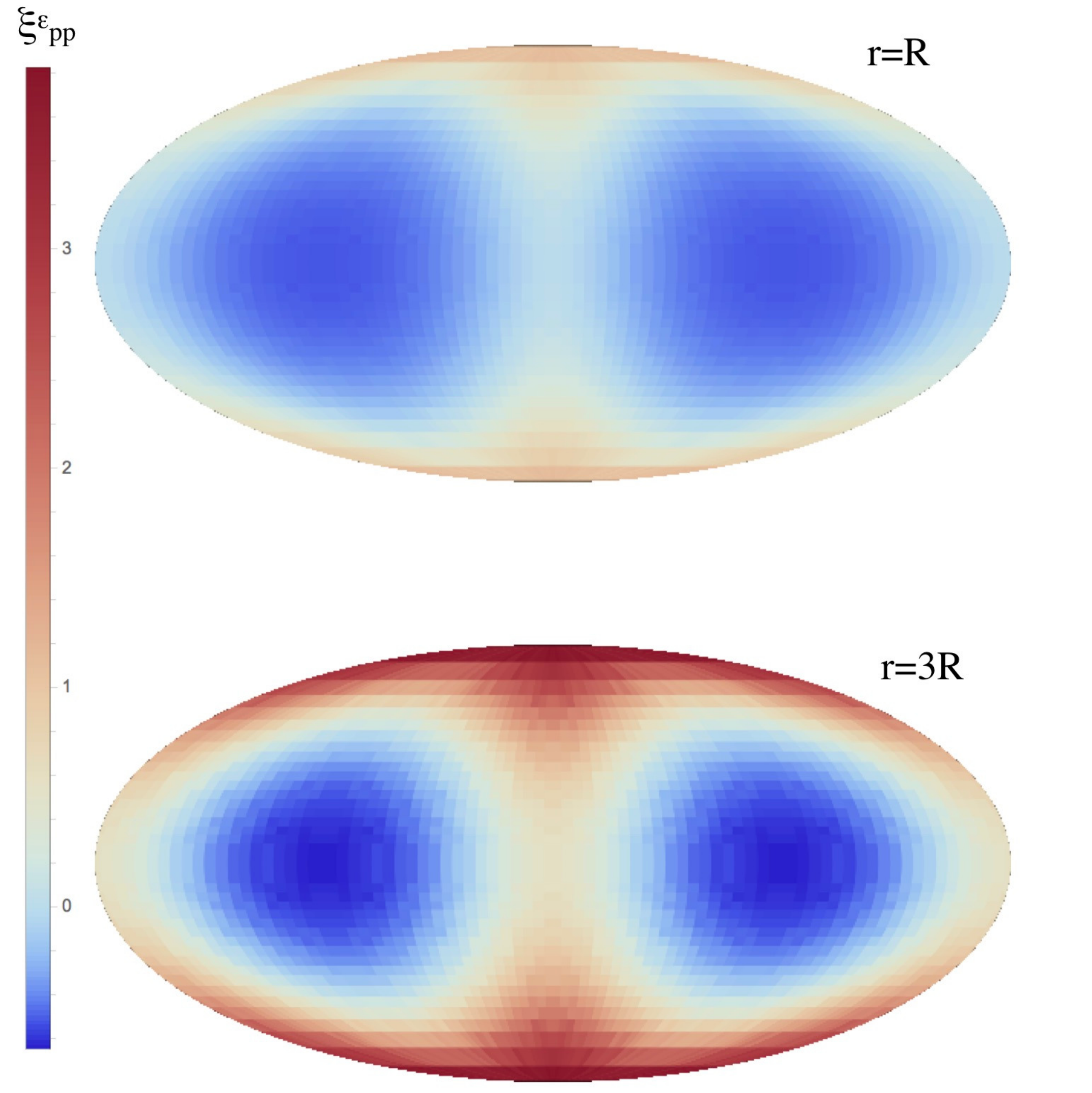}
	\includegraphics[width=\columnwidth]{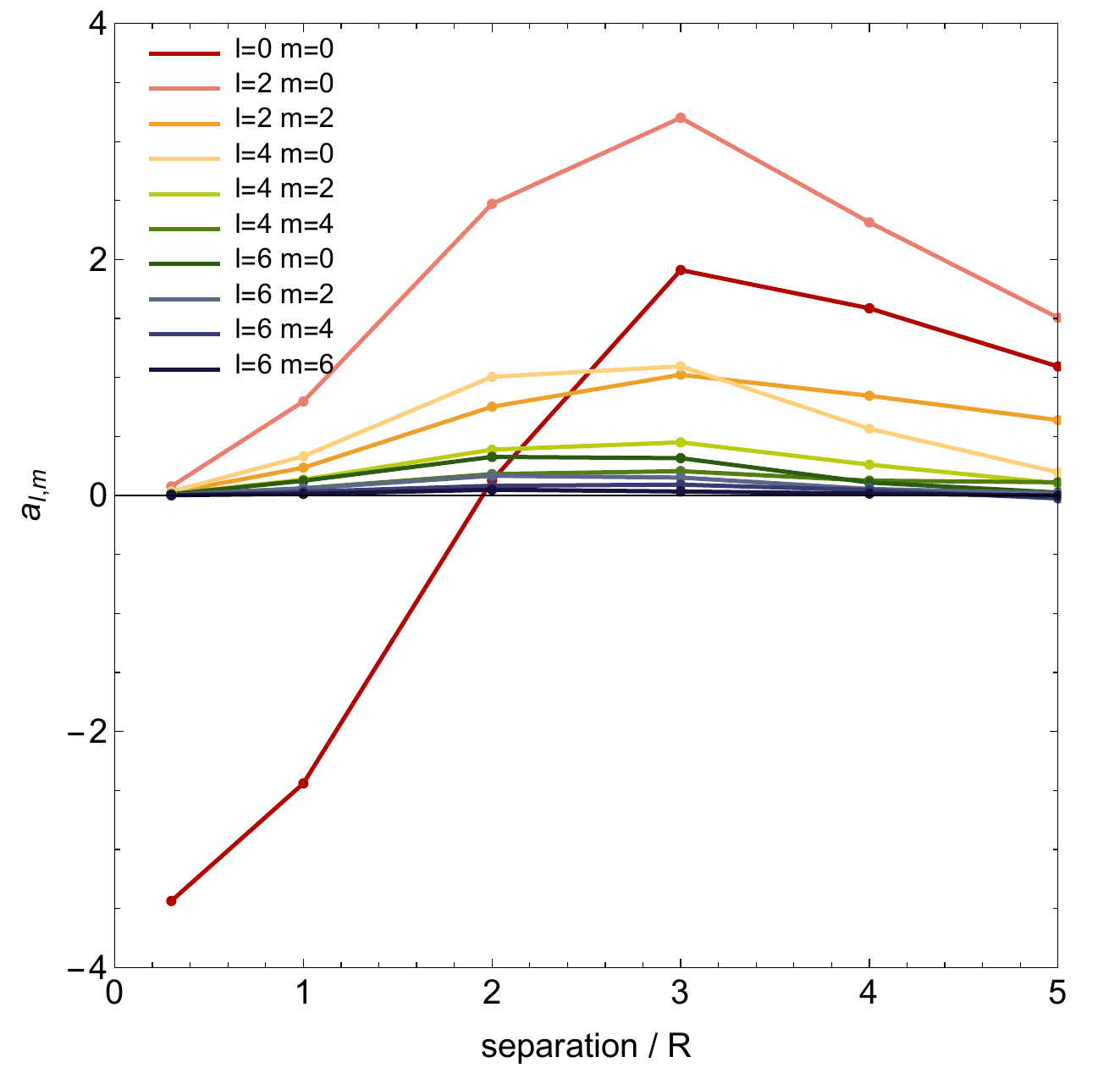}
	\caption{Oriented peak-peak correlation function for a Gaussian random field with spectral index $n_s=-2$. Left-hand panel: Mollweide projection of the sphere at $r=R$  (top panel) and $r=3R$ (bottom panel). The vertical axis is the major direction of the strain tensor, the minor direction corresponds to the two minima of the maps in dark blue while the intermediate direction corresponds to the saddle points at the centre of the maps and on the left and right edges. Right-hand panel: Decomposition of the oriented peak-peak correlation function onto spherical harmonics as a function of the separation.}
	\label{fig:pk-alm}
\end{figure*}

Because of the constraint on the sign of the second derivatives, the peak two-point correlation cannot be computed analytically.
Hence, 
we define the conditional probability that $x_{ij}$ and $x'_{ij}$ 
satisfy the PDF, subject to the condition that $x_i=x'_i=0$ and $x=x'=\nu$ and resort to Monte-Carlo 
methods in {\small MATHEMATICA} in order to evaluate numerically equation~(\ref{eq:xicrit}). Namely, we draw random 
numbers of dimension 
12 from the conditional probability that $x_{ij}$ and $x'_{ij}$ satisfy 
the PDF, subject to the condition that $x_i=x'_i=0$ and $x=x'=\nu$ (using {\tt RandomVariate}). 
For each draw $^{(k)}$ if $\lambda_3(x^{(k)}_{ij})<0 $ and  $\lambda_3(x'^{(k)}_{ij})<0 $, we keep the sample and evaluate $ {\rm det}(x^{(k)}_{ij})   {\rm det}(x'^{(k)}_{ij}) $ 
and otherwise we drop it. Eventually, 
\begin{multline}
\big\langle n_{\rm p}(\boldsymbol{X})n_{\rm p}(\boldsymbol{X'})\big\rangle
 \approx \frac{1}{N}
\sum_{k\in {\cal S}} \left[    {\rm det}(x^{(k)}_{ij})  {\rm det}(x'^{(k)}_{ij})\right]\\
\times {\cal P}(x=x'=\nu,x_{i}=x'_{i}=0) \,,
\end{multline}
where $N$ is the total number of draws, and $\cal S$ is the subset of the indices of draws 
satisfying the constraints on the eigenvalues.
The same procedure can be applied to evaluate the denominator
$\left\langle n_{\rm p}(\boldsymbol{X}) \right\rangle$. 
Equation~(\ref{eq:xicrit}) then yields  $\xi_{\rm pp}(r,\nu)$. This algorithm is embarrassingly 
parallel. It is straightforward to generalize it, for instance, to the computation of the correlation 
function $\xi_{\rm pp}(r,>\nu)$ of peaks {\sl above} a given threshold in density, or within a bin of rarity. 
Obviously, in those cases, the required number of draws 
is larger and increases with the value of the threshold (as the event $x>\nu$ becomes rarer) or the size of the  bin.

\subsection{Oriented stacks of peaks}

Going one step further, the idea is now to also do the calculation in the frame of the tidal tensor $\phi_{ij}$, which can be done by adding the constraint
\begin{multline}
{\cal B}_{\Epsilon}(\boldsymbol{X})=2\pi^2(\phi_{33}-\phi_{22})(\phi_{22}-\phi_{11})(\phi_{33}-\phi_{11})\delta_{\rm D}(\phi_{12})\\
\times\delta_{\rm D}(\phi_{13})\delta_{\rm D}(\phi_{23})
\Theta_{\rm H}(\phi_{33}-\phi_{22})\Theta_{\rm H}(\phi_{22}-\phi_{11}),
\end{multline}
which boils down to imposing the off-diagonal coefficients of the tidal tensor to be zero, the curvatures to be ordered and to adding the Jacobian of the transformation by means of the usual Vandermonde determinant $(\phi_{33}-\phi_{22})(\phi_{22}-\phi_{11})(\phi_{33}-\phi_{11})$ with an additional $2\pi^2$ due to the integration over the Euler angles.

As mentioned above, instead of using a $delta$-function at the mean $\langle \nu \vert R_c\rangle = \nu_{crit}$ we use a range of $\nu$ with spread $\Delta\nu$ about the mean. 
The oriented two-point correlation $\xi_{\rm pp}^{\Epsilon}(\ve{r},\nu,\Delta\nu)$ of peaks
in the range $\nu\pm \Delta\nu$ separated by $\ve{r}$ is then given by
\begin{equation}
\label{eq:oxicrit}
1+\xi_{\rm pp}^{\Epsilon}(\ve{r},\nu,\Delta\nu)=
 \frac{
\big\langle{\cal B}_{\Epsilon}(\boldsymbol{X})  N_{\rm p}(\boldsymbol{X})  N_{\rm p}(\boldsymbol{X'}) \big\rangle}
{\left\langle {\cal B}_{\Epsilon}(\boldsymbol{X})N_{\rm p}(\boldsymbol{X}) \right\rangle\left\langle N_{\rm p}(\boldsymbol{X}) \right\rangle}   \,,
\end{equation}
where the``localized'' density of peaks in the $\nu\pm \Delta\nu$ range is
\begin{multline}
\label{eq:oncrit}
N_{\rm p}(\boldsymbol{X})\!=\!-\!\left(\!\frac{\sigma_2}{\sigma_1}\!\right)^\text{3}\!
{\rm det}(x_{ij})\delta_{\rm D}(x_i)\Theta_{\rm H}(-\lambda_3)\\
\times\Theta_{\rm H}(x-\nu+\Delta\nu)\Theta_{\rm H}(\nu+\Delta\nu-x) \,.
\end{multline}
Once again we undertake a Monte Carlo evaluation of this correlation function. 
We define the conditional probability that $x_{ij}$, $\phi_{ii}$, $x'$ and $x'_{ij}$ 
satisfy the PDF, subject to the condition that $x_i=x'_i=0$ and $\phi_{i\neq j}=0$ and 
 use the Monte-Carlo 
methods in {\small MATHEMATICA} to determine  equation~(\ref{eq:xicrit}) numerically. To do this we draw random numbers of dimension 
16 from the conditional probability that $x_{ij}$, $\phi_{ii}$, $x'$ and $x'_{ij}$ satisfy 
the PDF, subject to the density peak conditions $x_i=x'_i=0$ and the $\phi_{i\neq j}=0$ strain eigenframe condition. For each draw $^{(k)}$ if $\lambda_3(x^{(k)}_{ij})<0 $, $\lambda_3(x'^{(k)}_{ij})<0 $, $\nu-\Delta\nu<x^{(k)}<\nu+\Delta\nu$, $\nu-\Delta\nu<x'^{(k)}<\nu+\Delta\nu$ and $\phi^{(k)}_{11}<\phi^{(k)}_{22}<\phi^{(k)}_{33}$,
 we keep the sample and evaluate $ {\rm det}(x^{(k)}_{ij})   {\rm det}(x'^{(k)}_{ij}) (\phi^{(k)}_{33}-\phi^{(k)}_{22})(\phi^{(k)}_{22}-\phi^{(k)}_{11})(\phi^{(k)}_{33}-\phi^{(k)}_{11})$;  
 otherwise we drop that draw. Eventually, 
\begin{multline}
\nonumber
\big\langle n_{\rm p}(\boldsymbol{X})n_{\rm p}(\boldsymbol{X'})\big\rangle
 \approx \frac{2\pi^2}{N}\left(\frac{\sigma_2}{\sigma_1}\right)^6{\cal P}(x_{i}=x'_{i}=\phi_{i\neq j}=0) \\
\times\sum_{k\in {\cal S}} 
   {\rm det}(x^{(k)}_{ij})  {\rm det}(x'^{(k)}_{ij})(\phi^{(k)}_{33}\!-\!\phi^{(k)}_{22})(\phi^{(k)}_{22}\!-\!\phi^{(k)}_{11})(\phi^{(k)}_{33}\!-\!\phi^{(k)}_{11}),
\end{multline}
where $N$ is the total number of draws, and $\cal S$ is the subset of the indices of draws 
satisfying the constraints on the eigenvalues.
The same procedure can be applied to evaluate the two terms of the denominator.
Equation~(\ref{eq:xicrit}) then yields  $\xi_{\rm pp}^{\Epsilon}(\ve{r},\nu,\Delta\nu)$. 

As an illustration, Fig.~\ref{fig:pk-pk-nu2} shows the resulting oriented peak-peak correlation function $\xi_{\rm pp}^{\Epsilon}(\ve{r},\nu,\Delta\nu)$ for $\nu=2$ and $\Delta\nu=0.5$ in the case of a power law power spectrum with spectral index $n_s=-2$. For simplicity, we only show here the slice corresponding to $y=0$ on the left-hand panel and the correlation function along the major (cyan) and minor (purple) axis on the right-hand panel. For comparison, we also compute the isotropic peak-peak correlation function and display it on the same plot with black dashed lines. In this case, we used 170 millions draws of the fields per spatial location. Interestingly, the result of this calculation is qualitatively quite similar to the oriented halo clustering in Lagrangian space obtained in Fig.~\ref{3D-DD}. Peaks cluster along the major axis of the strain tensor mostly between 3 and 4 smoothing lengths from the central object which is surrounded by an exclusion zone.
Although qualitatively correct, we cannot expect this simple model to be quantitatively correct. It depends significantly on the distribution of peak heights we have chosen, hence on the mass. For instance, as can be seen in Fig.~\ref{fig:pk-pk-nu}, high peaks tend to be more biased and more clustered. The size of the exclusion zone is also sensitive to the width of the height distribution. To go a step further we can cover the entire full mass range in the correlations by taking into account the specific rising forms of $\delta_{{\rm L}, {\rm crit}} (M)$ and the $M$-dependent spread about it with dropping $M$, as determined most accurately by fits to these distributions in the peak patch simulations, or with analytic approximations to such. Here the large mass asymptotic $\delta_{{\rm L}, {\rm crit}} (M)\approx 1.7$ was used. The $n_{\rm hpk}$ formula also has an extra ${\rm Tr}(x_{ij})$ multiplier, restricted to being of one sign because the piercing of the mass boundary first occurs from lower to higher $M$, an upcrossing constraint on the hierarchical peak patch trajectories. This is a familiar condition in the excursion-set treatments. An important extension is to use Top Hat rather than Gaussian filtering to better correspond to what the peak patch code does. Since the peak patch simulation is in fact a Monte Carlo method for computing quantities precisely, and it automatically take all effects into account, and more, it is better to just use the extremely large number of Lagrangian halo samples, possibly from multiple large boxes, to determined the correlations. Thus that is what we have done in the body of the paper. 

The analytic Monte Carlo formalism presented here also allows us to understand better the spherical harmonic decomposition of the 3D oriented peak-peak correlation function. This is illustrated on the right-hand panel of Fig.~\ref{fig:pk-alm} for a few separations. We also display on the left-hand panel the Mollweide projection of the oriented peak-peak correlation function at separation one and three smoothing lengths. The structure we find is quite similar to what was found for peak patch haloes in the main text. 
At small separation, the monopole is negative because there is an exclusion zone, and all the other harmonics are negligible there because the peak is locally close to isotropic (especially so for rare high mass peaks). Then there is an increase with increasing separation as the anisotropic cosmic web starts to play a role. The signal is largely dominated by the quadrupole as expected (due to the main filament) and it peaks at about 3 smoothing lengths -- which corresponds to the typical separation between the peaks -- before decreasing for large separations. Other harmonics up to $l=6$ appear but the signal clearly diminishes with the order of the harmonics. Directly below the $(2,0)$ harmonics comes the $(2,2)$ and $(4,0)$ terms which characterise the anisotropy due to the intermediate (i.e wall) direction and the bifurcation of filaments.
Following the isotropic case \citep[see][for a review]{
2018PhR...733....1D}, 
one could derive analytically the large separation bias expansion in this case. This is however beyond the scope of this paper.


\section{Spherical Harmonics}
\label{app:ylm}
\begin{figure*}
	\centering
    \includegraphics[width=\textwidth]{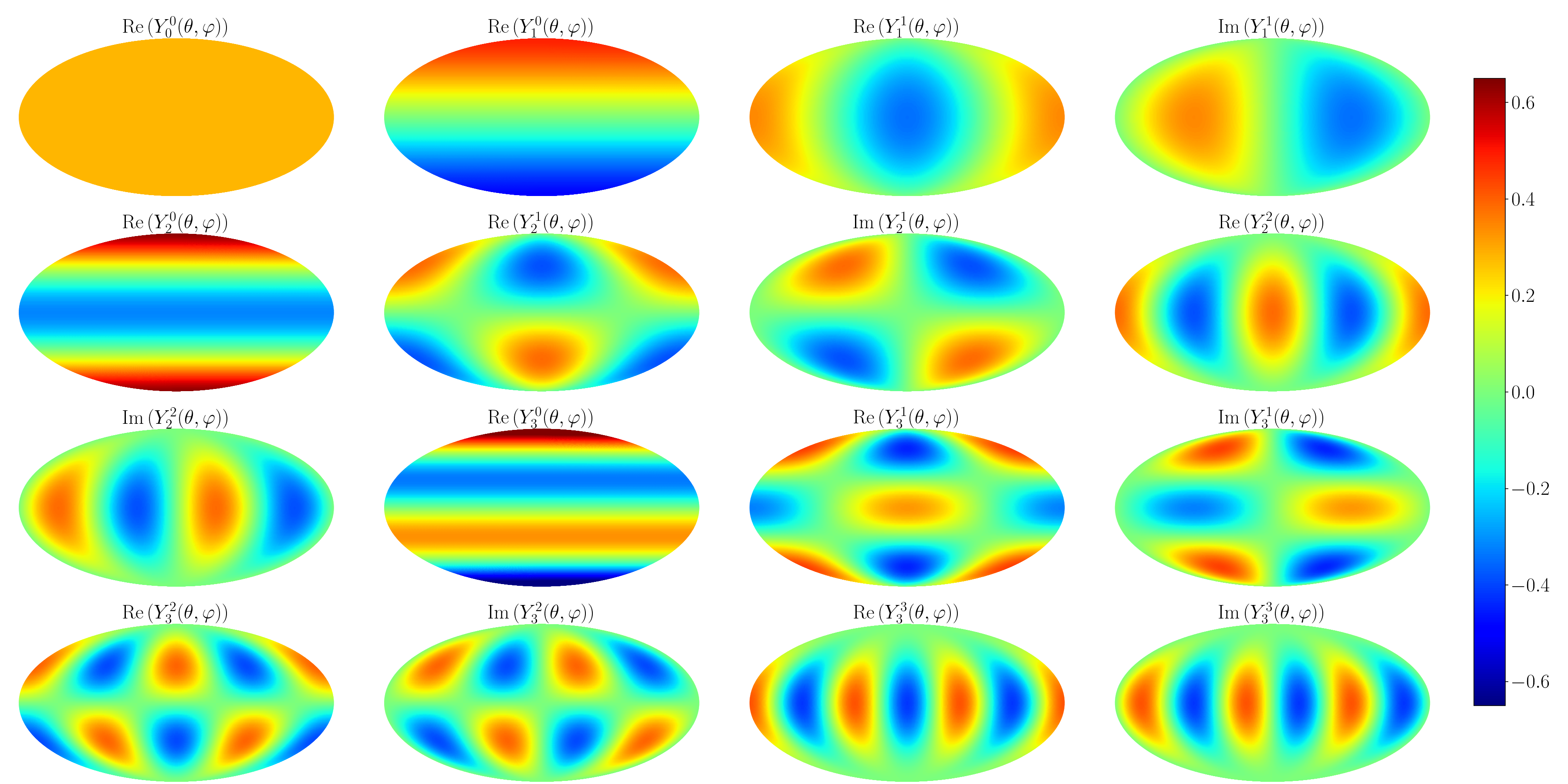}
	\caption{Representation of the first spherical harmonics, real and imaginary parts as labelled. $\theta = 0$ (respectively $\theta = \pi$) corresponds to the North (South) pole, $(\theta = \pi / 2, \varphi = 0)$ is at the center of the projections, and $\varphi$ values increase from left to right.}
	\label{ylm}
\end{figure*}

Fig.~\ref{ylm} shows Mollweide representations of the first spherical harmonics. The major direction is along the vertical axis, the intermediate direction crosses the centre and the left and right edges, and the minor direction passes through $(\theta=\pi/2, \varphi=\pm \pi/2)$ points.

\end{document}